\pdfoutput=1
\documentclass[10pt,preprint]{aastex}

\usepackage{natbib}
\usepackage{amssymb,amsbsy,amsmath}
\usepackage{graphicx} %for the \includegraphics command and figures put [final] before ...
\usepackage{ifthen,url}
\usepackage{verbatim}
\usepackage{booktabs}
\usepackage{multirow}
\usepackage{colortbl}
\usepackage{times}

\definecolor{mygrey}{rgb}{0.9,0.9,0.9}
\definecolor{gr}{rgb}{0.75,0.75,0.75}
\definecolor{lg}{rgb}{0.9,0.9,0.9}
\definecolor{dg}{rgb}{0.6,0.6,0.6}
\newcommand{\cclrgr}[0]{}
\newcommand{\cclrlg}[0]{}
\newcommand{\cclrdg}[0]{}
\newcommand{\notetoself}[1]{}

\shorttitle{Constraining the LISA Black Hole Mass Spectrum II}
\shortauthors{J. E. Plowman et al.}
\title{Constraining the Black Hole Mass Spectrum with LISA Observations II: Direct comparison of detailed models}
\author{Joseph E. Plowman\altaffilmark{1}, Ronald W. Hellings\altaffilmark{1}, Sachiko Tsuruta\altaffilmark{1}}
\altaffiltext{1}{Department of Physics, Montana State University, Bozeman, MT 59717}

\begin{document}

\bibliographystyle{apj}
%\include{aastex_journals}
%\maketitle

\begin{abstract}
A number of scenarios have been proposed for the origin of the supermassive black holes (SMBHs) that are found in the centres of most galaxies.  Many such scenarios predict a high-redshift population of massive black holes (MBHs), with masses in the range $10^2$ to $10^5$ times that of the Sun. When the Laser Interferometer Space Antenna (LISA) is finally operational, it is likely that it will detect on the order of $100$ of these MBH binaries as they merge. The differences between proposed population models produce appreciable effects in the portion of the population which is detectable by LISA, so it is likely that the LISA observations will allow us to place constraints on them. However, gravitational wave detectors such as LISA will not be able to detect all such mergers nor assign precise black hole parameters to the merger, due to weak gravitational wave signal strengths. This paper explores LISA's ability to distinguish between several MBH population models. In this way, we go beyond predicting a LISA observed population and consider the extent to which LISA observations could inform astrophysical modellers. The errors in LISA parameter estimation are applied with a direct method which generates random sample parameters for each source in a population realisation. We consider how the distinguishability varies depending on the choice of source parameters (1 or 2 parameters chosen from masses, redshift or spins) used to characterise the model distributions, with confidence levels determined by 1 and 2-dimensional tests based on the Kolmogorov-Smirnov test.
\end{abstract}

%\begin{keywords}
%black hole physics - early universe - gravitational waves - methods: statistical
%\end{keywords}


\section{Introduction}\label{chap:intro}
There is substantial evidence \citep[e.g.,][]{KormendyRichstone95, richstone1998sbh} for the existence of supermassive black holes (SMBHs) in the nuclei of most galaxies, the black hole in our own galaxy being the best studied and most clearly justified of these objects. However, the origin of these black holes remains an unsettled question. In one scenario, the more massive black holes formed from the merger and coalescence of smaller `seed' black holes created in the very early Universe \citep[e.g.,][]{Madau_Rees_2001}. Several models of this process have been proposed and numerically simulated \citep[e.g.,][]{KauffmannHaehnelt00,VHM03}. Typical seed black holes have masses $M_\bullet \sim 100M_\odot$ at high redshift (e.g., $z\sim$ 20), so these models predict an evolving population of massive black holes (MBHs), with masses that can cover the entire range from $\sim$ 100 to $10^9M_\odot$.

Younger members of this population fall into the intermediate-mass range ($100 M_{\odot}$ \dots $10^5 M_{\odot}$), and are not suited to electromagnetic detection, making it very difficult to verify a particular formation and evolution scenario or to discriminate between models. When the Laser Interferometer Space Antenna (LISA) is finally operational, however, it is likely that it will detect on the order of $100$ merging MBH Binaries in the range $M_{tot} \gtrsim 1000 M_{\odot}$. Since the differences between proposed population models produce appreciable effects in the subset of the population that LISA can detect, LISA observations should allow us to place constraints on the models. 

Once we have the LISA detected population in hand, we will need to determine how it constrains models of the astrophysical population which gave rise to it. Equivalently, we will want to know which model (with its associated parameters) is most likely to have produced the observed population. We also want to know how strongly the LISA data set will constrain the models: How dissimilar from the actual population does a model population need to be before it can be distinguished based on the LISA data? Fully answering this question requires considering LISA's ability to detect sources, the parameter estimation errors associated with sources inferred from the LISA data, and, eventually, the {\em a posteriori} source parameter distributions of sources extracted from the LISA data. It also requires considering how to quantify the differences between populations as they would be observed from the LISA data and associating a degree of confidence with these difference quantities.

In this paper, we apply LISA parameter estimation errors and detection thresholds to various population models, and compare the resulting distributions of estimated parameters. This work goes beyond the comparisons which we presented in \cite{plowetal2009I}, comparing detailed distributions predicted by several population models rather than roughly binned distributions gleaned from the literature. The comparisons are extended to two-dimensional comparisons of the distributions of estimated parameters, using modified versions of the one and two-dimensional Kolmogorov-Smirnov (K-S) tests which are sensitive to differences in event counts as well as to differences in the parameter distributions. We consider how the distinguishability of the models depends on a variety of factors such as the length of the observation window and the binary parameter dimensions being compared. The results demonstrate the ability of LISA to constrain the astrophysics of MBH formation, and shed light on the directions of future research in MBH population models, LISA instrument models, and model comparison efforts.

\section{BBH Parameters \& their Errors}\label{sec:BBHparams}
Each black hole binary may be characterised by a number of parameters, which are sometimes  divided into two categories. The first category is the {\em intrinsic} parameters which have to do with the local properties of the binary in its rest frame. They are $m_{1}$, the mass of the primary, $m_2$, the mass of the secondary, two sets of initial black hole spin and orientation parameters (relative to the orientation of the binary orbit), and either the initial orbital separation, $a$, frequency, $f$ (related to $a$ by Kepler's third law) or time to coalescence, $t_c$, (related to $a$ by the quadrupole formula)\footnote{In general, the list of intrinsic parameters also includes the binary's initial orbital eccentricity. Here, however, attention is restricted to the simplified case of circular orbits.}. A set of mass parameters equivalent to $m_1$ and $m_2$ but more directly related to the gravitational waveform, is the chirp mass, ${\cal M}_{c}=(m_1 m_2)^{3/5}(m_1+m_2)^{-1/5}$, and the symmetric reduced mass ratio $\eta=(m_1m_2)/(m_1+m_2)^2$. We use as spin parameters the dimensionless spin variables $\chi_1$ and $\chi_2$, related to the spin angular momenta $s_1$ and $s_2$ of the BHs according to $s_1 = \chi_1 m_1^2$ and $s_2 = \chi_2 m_2^2$. $\chi_1$ and $\chi_2$ are restricted to the range $0\dots 1$; this is discussed in, for instance, \cite{Hartle}.

The remaining parameters fall into the second category, and are called {\em extrinsic} parameters. These have to do with the binary's location and orientation with respect to the LISA constellation. They are the luminosity distance $D_{L}$ (or equivalently, the redshift $z$), the (initial) principle gravitational wave polarisation angle $\psi$, the (initial) binary inclination $\iota$, the sky location angles $\theta$ and $\phi$, and the initial phase of the binary orbit $\Phi_0$.

To first order, the mass dependence of the waveform is on ${\cal M}_c$ only. Other mass variables such as $\eta$ enter into the amplitude and phase of the waveform, along with the chirp mass, at higher orders. The luminosity distance affects the overall amplitude of the gravitational wave, while the orbital inclination changes the relative amplitudes of the $h_+$ and $h_\times$ components of the metric. The sky position variables produce a modulation of the wave amplitude and frequency as the LISA spacecraft orbit the sun, changing the relative orientation of the LISA constellation and the plane of the GW as well as the velocity of the detector with respect to the source. To second Post-Newtonian order, the BH spins affect the phase evolution of the orbit and the orbital orientation due to spin-orbit and spin-spin interactions.

%\begin{figure*}
%   \centering
%   \includegraphics[width=1\textwidth]{figures/kerr_waveform_plot.pdf}
%   \caption{Example spinning binary waveform. The binary mass ratio is $m_1/m_2 = 10$ and the (redshifted) ${\cal M}_c = 10^6$. The spin of the more massive binary begins aligned with the orbit, while the spin of the smaller binary is rotated by $40^\circ$ with respect to the orbit. The waveform has a Hann window applied to the final orbits, to reduce ringing in the Fourier domain.}
%   \label{fig:waveformexample}
%\end{figure*}

Redshift and luminosity distance are used interchangeably as source parameters, with the relationship between them determined from the standard WMAP cosmology ($\Omega_M=0.27$, $\Omega_\mathrm{vac}=0.73$, $\Omega_\mathrm{rad}=0.0$, and a Hubble constant of 71 km/s/Mpc). Gravitational lensing can impart an error in the determination of the luminosity distance of order $\sim 10\%$ at $z=15$ \citep[see][]{HolzLinder05}, but we have not incorporated this effect into our analysis.

\begin{figure*}
   \centering
   \includegraphics[width=1\textwidth]{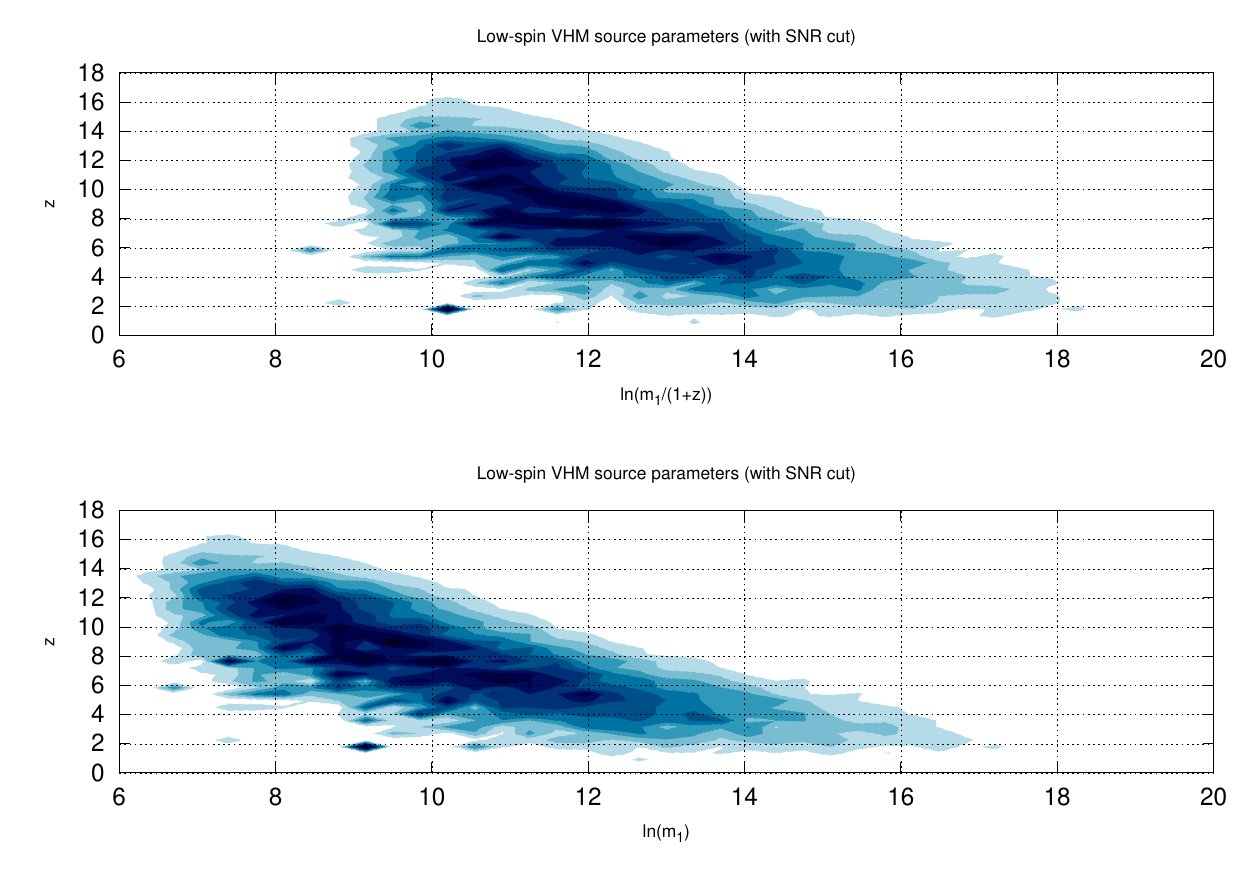}
   \caption{Example plot showing the difference between parameter distributions expressed in terms of redshifted (top) and non-redshifted (bottom) mass variables. Note that an SNR cut has been applied to these distributions.}
   \label{fig:redshiftedmassexample}
\end{figure*}

In the gravitational wave signal, the masses are scaled by the redshift, so that the natural mass variables for gravitational wave data analysis are redshifted. For instance, the redshifted chirp mass is ${\cal M}_c (1+z)$, although the reduced mass ratio remains unchanged since it is dimensionless. Since the luminosity distance (and therefore the redshift, assuming a particular cosmology) is poorly determined from the gravitational wave signal and the redshifted masses are very well determined, the natural mass variables for GW population analyses are redshifted. We therefore compare model astrophysical source distributions as functions of the redshifted masses, rather than expressing them as functions of rest-frame masses, which is more customary (see Figure \ref{fig:redshiftedmassexample}). Mass variables used in this paper are redshifted unless otherwise noted.

Since the predictions of the population studies are given as functions of masses and redshift, the binary parameters are best divided into two sets in a {\em different} way, for purposes of constraining the population models. The first set, consisting of ${\cal M}_{c}$, $\eta$, the two spin magnitude parameters $\chi_1$ and $\chi_2$, and $D_L$ (or $z$), are what we will call \emph{population} parameters, since these are the parameters that characterise the population model predictions. The remaining parameters, $\psi$, $\iota$, $\theta$, $\phi$, $t_c$, and $\Phi_0$, represent particular samples drawn from the population model and will be referred to as \emph{sample} parameters. Sample parameters have distributions that are essentially stochastic and contain no useful information about the astrophysical processes which give rise to the black hole population. The model results we employ do not give spin orientation distributions, so we also assume that they are each independently stochastic (uniformly distributed on the sky) and list them as sample parameters, although this may not be true in general \citep[see][]{BertiVolonteri08}. Table \ref{parmtable} summarises parameters and their classifications.

\begin{table}
\caption{The parameters which define a source. There are a variety of combinations of $m_1$ and $m_2$ which can be used to specify the masses of the source, depending on context.}\label{parmtable}
\centering
\begin{tabular}[!h]{c|p{0.6\textwidth}|c}
%\hline 
Name & Description & Classification \\
\hline
\hline
$D_L$ & The luminosity distance & Population \\
\hline
$m_1$ & The mass of BH \# 1 & Population \\
\hline
$m_2$ & The mass of BH \# 2 & Population \\
\hline
$\chi_1$ & The dimensionless spin parameter of BH \#1 & Population \\
\hline
$\chi_2$ & The dimensionless spin parameter of BH \#2 & Population \\
\hline
$e$ & The eccentricity of the binary (assumed to be zero in this work) & Population \\
\hline
$\theta$ & Co-latitude angle. & Sample \\
\hline
$\phi$ & Azimuthal location. & Sample \\
\hline
$t_c$ & Initial time to coalesce & Sample \\
\hline
$i$ & The (initial) orbital inclination angle & Sample \\
\hline
$\Psi$ & The (initial) rotation angle of the plane of the orbit & Sample \\
\hline
$\phi_{\textrm{\tiny GW}}$ & The (initial) gravitational wave phase & Sample \\
\hline
$\theta_{s1}$, $\phi_{s1}$ & The spin orientation parameters of BH \#1 & Sample \\
\hline
$\theta_{s2}$, $\phi_{s2}$ & The spin orientation parameters of BH \#2 & Sample
%\hline
\end{tabular}

\end{table}

Despite the fact that the sample parameters are not part of the
intrinsic astrophysical model, they can have a dramatic effect on
LISA's source characterisation capabilities. This is because population parameters can be correlated with sample parameters and all parameters
must be fit to the data in the process of extracting the
population parameters of interest. In particular, the luminosity distance is correlated with the inclination angle and orbital phase in such a way that the luminosity distance error varies significantly depending on the source's inclination. It is therefore important to incorporate the effects of variation in the sample parameters on the errors in the source parameters. 

Parameter error estimation for black hole binaries detected via
gravitational wave emission has been discussed by many researchers
\citep{CutlerFlanagan94, Vallisneri08, MooreHellings02,
CrowderThesis06}. The parameter estimation errors used in this work are based on the Fisher
information matrix, a standard technique used in many of the papers listed above. It is important to remember, however that the Fisher error estimate is
accurate only when the parameter uncertainties are small compared
to the characteristic scales of the system being fit
\citep{Vallisneri08}, a condition that is not well satisfied for
all of the binaries being modelled here.

Rather than write our own Fisher error estimation codes, we have
made use of the recent Montana/MIT group code, discussed in \cite{LangHughes09} and in \cite{LISAPE2009}, which was provided to us by Neil Cornish and Scott Hughes. This waveforms used by this code \citep[based on][]{Apostolatosetal94} incorporate the effects of higher harmonics in the wave form, spins and spin precession, but it executes rather slowly \citep[recent developments which can result in a substantial speedup are described in][]{CornishFF2010}. It is accurate to second Post-Newtonian order.

The codes take as input a set of source parameter
values, $\lambda_j$, and output a set of standard deviations, $\sigma_i$, for the detected
parameter values, $\hat\lambda_i$. Each detected parameter,
$\hat\lambda_i$, is assumed to have a Gaussian probability density
with mean $\lambda_i$ and standard deviation $\sigma_i$,
\begin{equation}
   p_{D}(\hat\lambda_{i}|\lambda_i)=\frac{1}{\sqrt{2\pi\sigma_{i}^{2}}}
     \exp{\Bigg[
   {-\frac{\big[\hat\lambda_{i}-\lambda_{i}\big]^{2}}{2\sigma_{i}^{2}}}\Bigg]}.
   \
   \label{LCpdf}
\end{equation}

Note that equation \ref{LCpdf} neglects the (sometimes large) correlations between the parameter estimation errors, effectively assuming that the axes of the error ellipse are parallel to the parameter directions. In cases where the parameter estimation errors are strongly correlated, this exaggerates the effects of the errors. In Section \ref{chap:2D_direct_comparisons}, however, we find that only one parameter (namely, $\chi_2$) has errors which significantly effect the estimated parameter distributions predicted by the models, so this simplification should not affect the results presented here\footnote{If we were using rest-frame masses rather than redshifted masses, on the other hand, we could not ignore this effect since the correlation between (rest-frame) mass and distance is quite significant}. Also note that we are only considering parameter estimation errors due to noise in the LISA instrument; other sources of noise, such as variations in the amplitudes due to weak gravitational lensing effects, are not included.

In \cite{plowetal2009I}, the parameter estimation errors were applied using LISA error distributions which were averaged over a Monte Carlo ensemble of many sources, each having randomly-chosen values for the sample parameters. Such error distributions are referred to as `error kernels'. This paper uses a second method, described in more detail in Section \ref{chap:2D_direct_comparisons}, where a new set of random sample parameters is generated for every source in a realisation of a population. It has the advantage, compared to the error kernel, that it is easier to implement and the up-front computational cost is quite low. It also scales easily to the case where there are more than a few population parameters, while the error kernel does not. On the other hand, once error kernels are computed, new realisations of a LISA detected parameter set can be produced very quickly. The error kernel also provides a convenient way to visualise LISA's average parameter determination error.

\section{Model Comparison}\label{chap:modelcomp}

The tests which we use to compare distributions of LISA estimated parameters are based on the Kolmogorov-Smirnov (K-S) test. Its nonparametric and unbinned nature makes it well suited to the sparsely distributed ($\sim 100$ sources spread over multiple decades and parameter dimensions) MBH population, unlike the binned $\chi^2$ test. One drawback of this test is that in practice, it is restricted to comparing two-dimensional distributions.

\begin{figure*}
   \centering
   \includegraphics[width=1\textwidth]{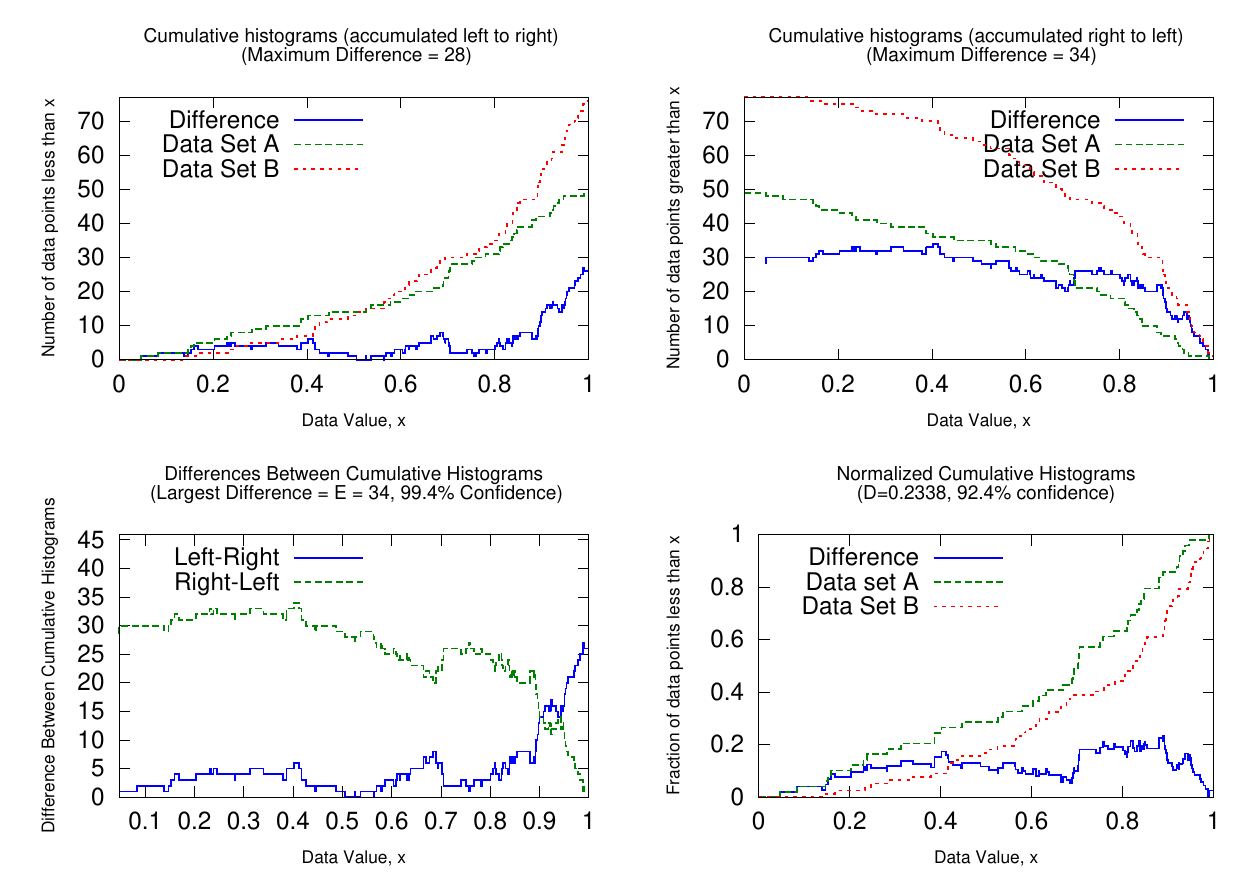}
   \caption{Comparison of two example data sets illustrating the calculation of the $E$ statistic. Top plots show the two cumulative histograms used in calculating the $E$ statistic, as well as the differences between them. The lower left plot shows only the differences between the two pairs of histograms, the largest of which is the $E$ statistic. Shown in the lower right are the normalised cumulative distribution functions on which the K-S test is based. These synthetic data sets were created with the probability density function (PDF) $p(x) = x^\alpha/(1+\alpha)$ (x ranging from zero to one), and Poisson distributed total number counts. Data set A had $\alpha = 0.6$ and Poisson parameter 56, while data set B had $1.5$ and 72.}
   \label{fig:ksedfs}
\end{figure*}

In its one-dimensional form, the K-S statistic is almost deceptively simple to implement. First, form a cumulative histogram from each data set, and normalise both to one. These are an estimate of the Cumulative Distribution Functions (CDFs) from which the data sets were drawn, and are sometimes referred to as `Empirical Distribution Functions' (EDFs); Figure \ref{fig:ksedfs} shows example EDFs and cumulative histograms of two data sets. The K-S statistic, $D$, is the maximum difference between the two EDFs. Because the size of this difference does not change under reparametrisation of the $x$ axis (i.e., arbitrary stretching or shrinking of the separation between data points), the K-S test is nonparametric. For instance, the same values of the statistic are obtained for comparison of the data sets $\{A_i^2\}$ and $\{B_i^2\}$, and for the data sets $\{A_i\}$ and $\{B_i\}$ (assuming all of the data points are positive). For the standard one-dimensional K-S test, the confidence level is given by \cite{nr} as
\begin{equation}\label{KS1conf}
Q_{KS}\Big(\big[\sqrt{N_e} + 0.12 + 0.11/\sqrt{N_e}\big] D\Big),
\end{equation}
where the function $Q_{KS}$ can be evaluated as the series
\begin{equation}\label{qks}
Q_{KS}(\lambda) = 2\sum_{j=1}^{\infty} (-1)^{j-1}e^{-2j^2\lambda^2}
\end{equation}
and the `effective' number of data points, $N_e$, is determined by $N_A$ and $N_B$, the numbers of points in data sets $A$ and $B$:
\begin{equation}\label{ne1dks}
N_e = \frac{N_AN_B}{N_A+N_B}.
\end{equation}
 
This standard K-S test is sensitive only to variation in the CDFs of the two populations from which the data sets are drawn. As long as the EDFs of the two data sets are similar, the test will report that the null hypothesis is true, even if the number of counts in the two sets are very different. Since the number of detected LISA sources is also an important prediction of a population model, we have constructed a modified test that is also sensitive to differences in both the CDFs and the number of sources. 

We begin by defining a new statistic, $E$, which differs from the K-S $D$ statistic most notably in that the cumulative histograms going into its calculation are not normalised. Instead of ranging from $0$ to $1$, values of the statistic will range from $0$ to $N_A$ and $0$ to $N_B$. Thus, significant differences in the total count rate can also lead to significant differences in the statistic. We have found that using only the standard cumulative histogram, which accumulates data counts going from left to right (low to high data values), is sensitive primarily to distributions whose differences occur at the upper end of the distribution, and insensitive to differences on the low end (the upper two panels of Figure \ref{fig:ksedfs} illustrates this). In order to allow differences between the model CDFs more opportunities to manifest themselves, we consider two sets of differences between the data EDFs, one set with EDFs accumulated ranging from small values of the data variable to large values, and the other set ranging from large values to small values (see Figure \ref{fig:ksedfs}). The statistic itself is taken to be the largest difference found in the sets. Note that this statistic retains the useful nonparametric nature of the standard K-S test. Then, we modify the null hypothesis:

\begin{quotation}
Suppose that the two data sets are drawn from the same CDF, {\em and} that the numbers of data points (i.e., the total number of sources) are the result of a Poisson process with the same rate parameter. With what {\em frequency}, $p$, would one obtain values of the $E$ statistic at least as extreme as that which we actually observed?
\end{quotation}

\begin{figure*}
   \centering
   \includegraphics[width=1\textwidth]{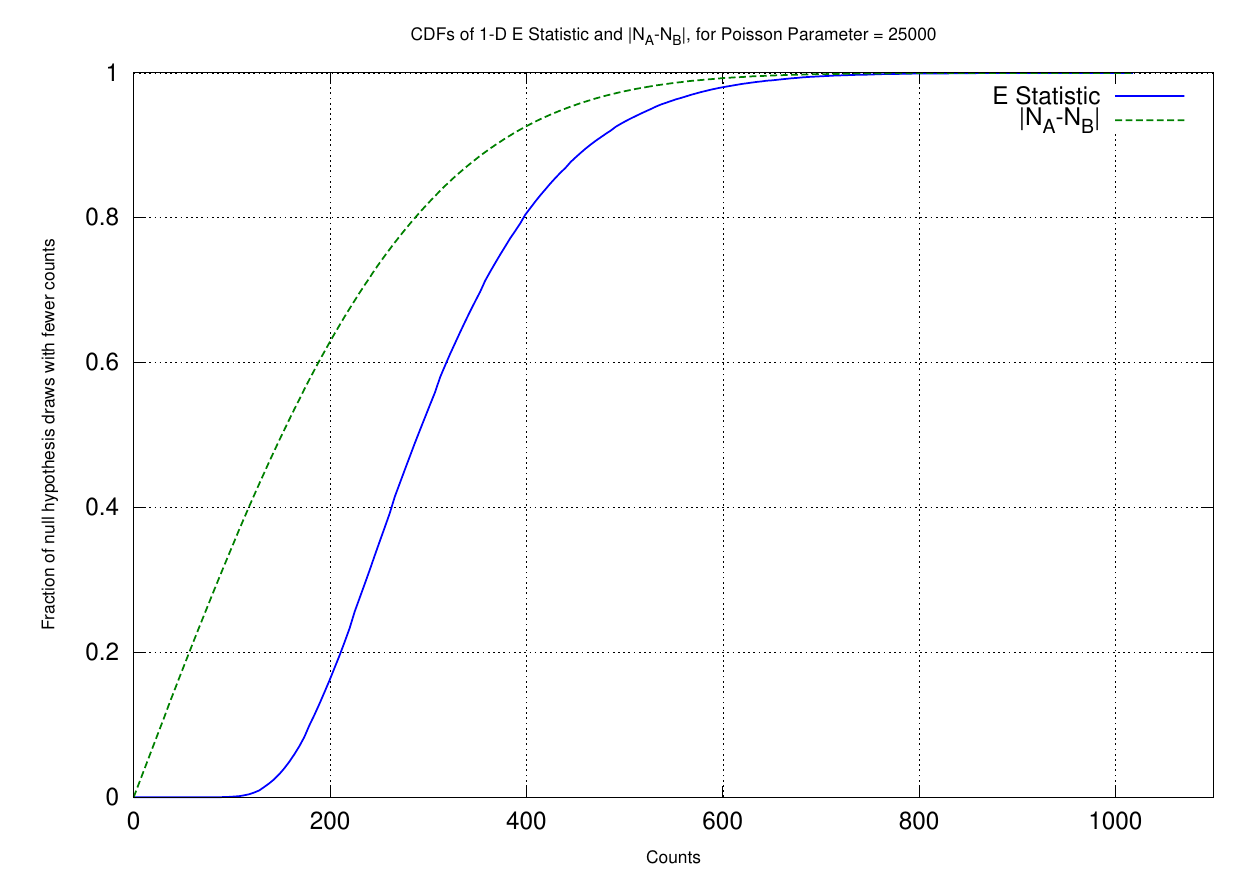}
   \caption{Cumulative distribution function of one-dimensional $E$ statistics, for $2.5\times 10^5$ draws given the null hypothesis. For each draw, two simulated data sets were produced, each with total number of data points drawn from a Poisson distribution with $\lambda = 25000$ and individual data points drawn from a uniform distribution over $0\dots 1$. The $E$ statistics resulting from these draws were then tabulated into a cumulative histogram and normalised, resulting in this plot. Shown for comparison is the equivalent CDF of the differences between the number counts, $|N_A-N_B|$.}
   \label{fig:Ecdf1D}
\end{figure*}

Rather than calculate the CDF of the test statistic analytically, we calculate it numerically using Monte Carlo draws given this null hypothesis. Since the statistic is nonparametric, our null hypothesis draws can use data points that are uniformly distributed without loss of generality. Moreover, we have numerically determined that the mean of the (one-dimensional) $E$ statistic (in the case of the null hypothesis) scales with the Poisson rate parameter, $\lambda$, approximately as
\begin{equation}\label{emean1D}
\langle{E}\rangle = 1.773\sqrt{\lambda} - 0.487,
\end{equation}
and its variance scales approximately as
\begin{equation}\label{evar1D}
\textrm{Var}(E) = \frac{1}{2}(1.046\lambda + 0.5).
\end{equation}
Thus, we can use a fiducial CDF calculated for one $\lambda$ to find the CDF for any other by scaling $E$ according to the above equations. Note, however, that the statistic can only take on discrete values, so CDFs calculated with small $\lambda$ will not smoothly cover the range of $E$. In the case of two data sets, the average of the two number counts, $(N_A+N_B)/2$, should be used as an estimator of $\lambda$. Figure \ref{fig:Ecdf1D} shows a fiducial CDF calculated using $2.5\times 10^5$ null hypothesis draws with $\lambda=2.5\times 10^4$, which should be sufficient for most uses.

\begin{figure*}
   \centering
   \includegraphics[width=1\textwidth]{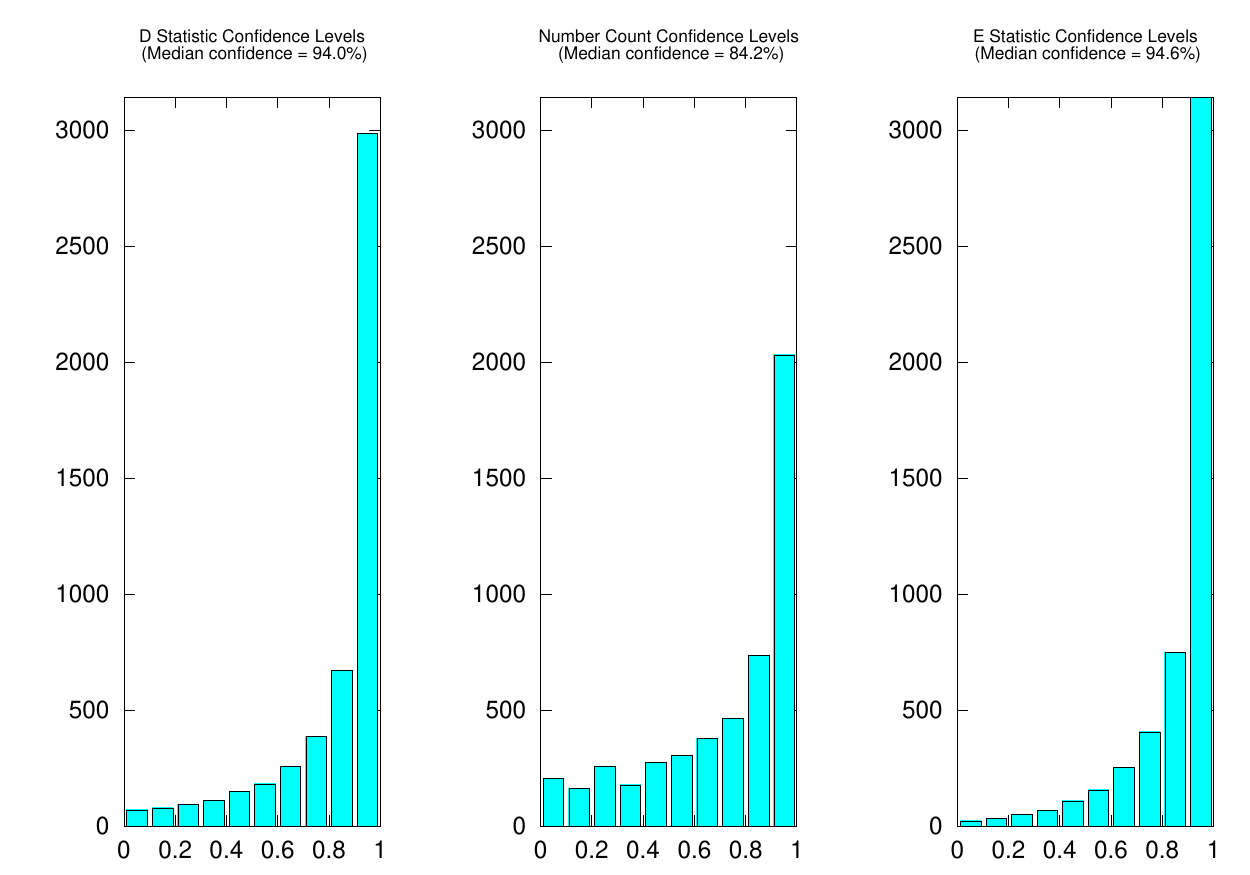}
   \caption{Example confidence levels for the K-S test (left), the $E$ statistic (right), and the statistic formed by taking the difference between the total number counts (a $\chi^2$ statistic with only one degree of freedom). These synthetic data sets were produced in the same way as in Figure \ref{fig:ksedfs}.}
   \label{fig:example_confs}
\end{figure*}

The differences between the various tests are illustrated by the confidence levels in Figure \ref{fig:example_confs}.

\subsection{Two-Dimensional Tests}\label{sec:2DEtest}

Since each detected LISA source has a set of best-fit parameters (masses, redshifts, spins, etc) rather than a single parameter, the model comparison tests employed should be sensitive in multiple dimensions as well. In this work therefore, we also employ a version of the two-dimensional K-S test, modified to be sensitive to differences in the total event counts. 

The two-dimensional test statistics are somewhat more complicated than those for the one-dimensional case, because the CDF and the cumulative histogram are not well defined in two or more dimensions (there's a well defined rank ordering in one dimension, but not in two or more). However, a working surrogate can be obtained in two dimensions \citep{Peacock83}. For each data point $(x_{Ai},y_{Ai})$ in the set $A$, find the fractions (one for data set B, one for data set A) of data points in the four quadrants defined by $(x < x_{Ai}, y < y_{Ai})$, $(x < x_{Ai}, y > y_{Ai})$, $(x > x_{Ai}, y < y_{Ai})$, and $(x > x_{Ai}, y > y_{Ai})$, for data set A and for data set B. Then, find the point $x,y$ and choice of quadrant for which the difference between these fractions is the greatest. Do the same thing, but running over each data point in B, rather than each data point in A, and average the two fractions\footnote{\cite{nr} give an algorithmic definition in the C language, which may be more clear}. This is the two-dimensional $D$ statistic, first implemented in this fashion by \cite{FasanoFranceschini87}. The two-dimensional $E$ statistic is the same except that, in place of the fraction of data points in each quadrant, we use the number of data points in each quadrant.

In the case of the null hypothesis, the distribution of the $D$ statistic is very nearly independent of the shape of the two-dimensional distribution of the data points \citep{nr}, and this appears to be the case for the $E$ statistic as well (see Section \ref{sec:testvalidation}). The confidence levels (i.e., the CDFs for draws given the null hypothesis) for both $D$ and $E$ can be calculated using draws given the null hypothesis, as before. For the $D$ statistic, \cite{nr} give an equation (their eq. 14.7.1) that can be used to approximate the CDF.

\begin{figure*}
   \centering
   \includegraphics[width=1\textwidth]{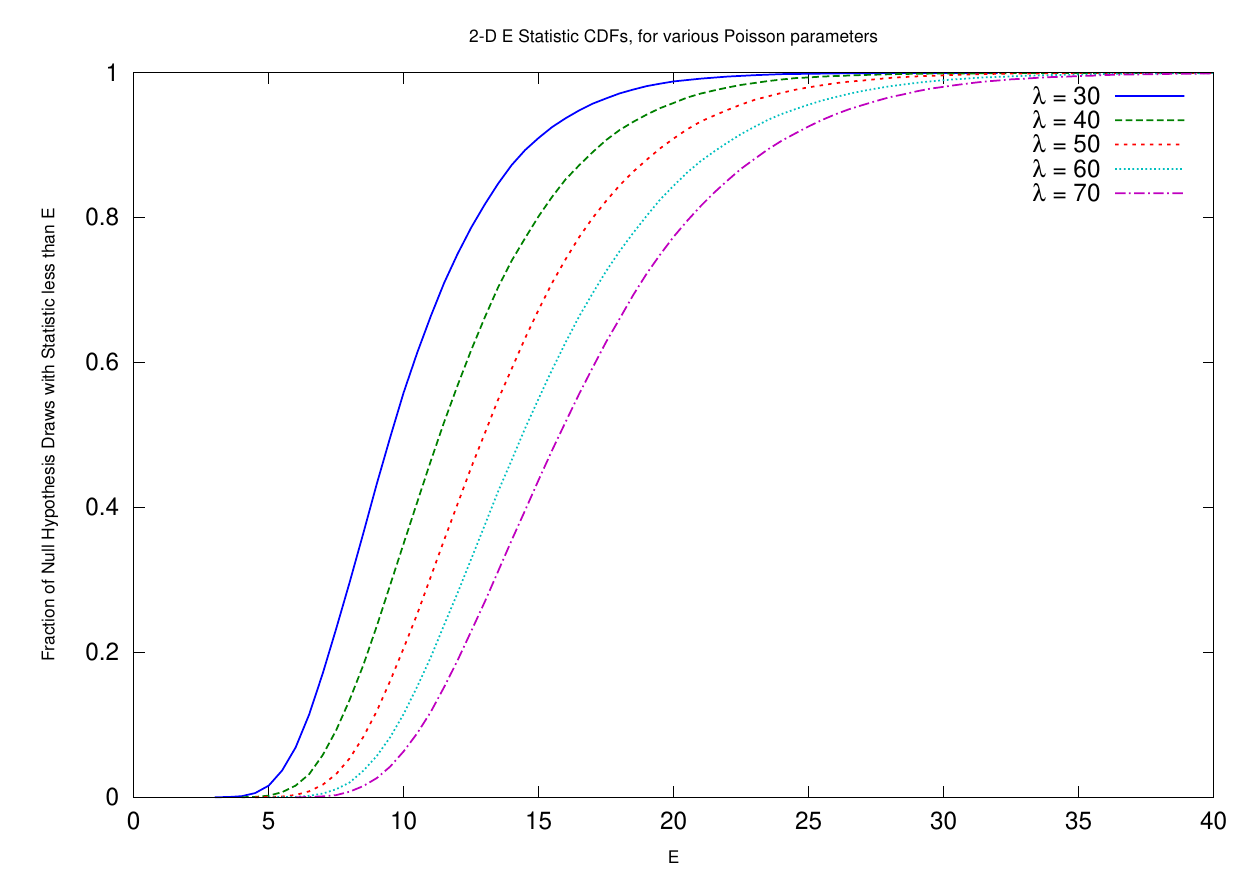}
   \caption{Cumulative Distribution Functions for two-dimensional $E$ statistics, each made using $5\times 10^4$ draws given the null hypothesis. Calculated in a similar fashion to that described in the caption of Figure \ref{fig:Ecdf1D}.}
   \label{fig:example_2D_ECDFs}
\end{figure*}

As with the one-dimensional case (e.g., Figure \ref{fig:Ecdf1D}), $E$ statistic CDFs were determined by Monte Carlo draws consistent with the null hypothesis, rather than an analytical approximation formula. We have not attempted to estimate scaling rules similar to equations \ref{emean1D} and \ref{evar1D} for the two-dimensional $E$ statistic, however. Instead, we calculate the CDF directly using Monte Carlo draws for each sample size, and store the results for future reference. Figure \ref{fig:example_2D_ECDFs} show several such CDFs.

\subsection{Validation of Tests}\label{sec:testvalidation}

Before we apply these tests, we would like to know if the confidence levels we obtained in the previous sections are, in fact, valid for any distribution. We have therefore performed some test cases to see if the distribution of confidence levels (in the case of the null hypothesis) has any appreciable dependence on the distribution used to make the draws. Recall that the confidence level for some value of a statistic is the percentage of draws given the null hypothesis which are less than or equal to that value. If the null hypothesis is true, therefore, the confidence levels resulting from a series of data sets drawn from the population will approximate\footnote{Since the statistics in question take on discrete values, the confidence levels will also take on discrete values. Thus, a cumulative histogram of the confidence levels will increase in discrete steps rather than the uniform distribution's straight line of unit slope.} a uniform distribution: 10\% of null hypothesis draws will have confidence levels less than 10\%, 20\% will have draws less than 20\%, etc. We can use this fact to check our method of calculating the confidence levels by performing draws given the null hypothesis (that the data sets both have the same CDF and number counts drawn from the same Poisson distribution) which use different CDFs than those used in the calculation of the confidence levels. If we obtain a uniform distribution of confidence levels for data sets which have distributions significantly different than those used in the construction of the confidence levels, then it is likely that our method of calculating confidence levels is correct in general. 

\begin{figure*}
   \centering
   \includegraphics[width=1\textwidth]{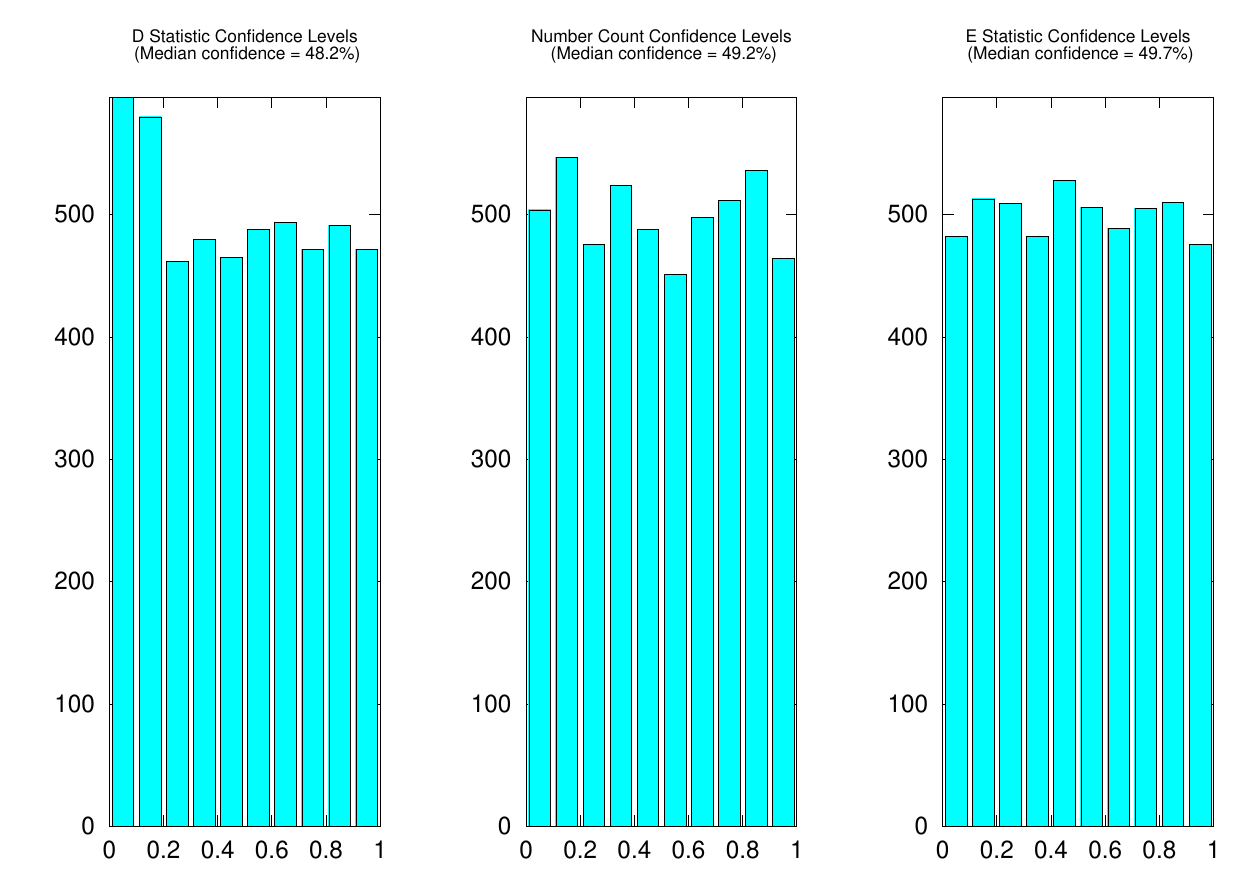}
   \caption{Histogram of confidence levels comparing data sets drawn given the null hypothesis. Individual data points were Lorentzian distributed, and number counts varied according to a Poisson distribution. Poisson rate parameter and Lorentzian median and HWHM parameters varied randomly from one comparison to the next, as discussed in the text.}
   \label{fig:etest_1D_test_histogram}
\end{figure*}

For the one dimensional tests, we have used as a test distribution a Lorentzian, or Cauchy distribution, which has the probability density function
\begin{equation}\label{lorentzianpdf}
p(x) = \frac{1}{\pi}\frac{\gamma^2}{(x-x_0)^2+\gamma^2},
\end{equation}
where $x_0$ is the median of the Lorentzian, and $\gamma$ is its half-width at half maximum (HWHM). The Lorentzian distribution is semi-pathological because it has very large tails, so large that its mean and standard deviation are undefined (although it is, of course, normalisable), and it consequently makes for an interesting test case. Rather than making all pairs of null hypothesis draws from identical distributions, we have chosen a different set of random distribution parameters for each pair of synthetic data sets drawn given the null hypothesis (naturally, each comparison performed was between data sets with the same set of distribution parameters). The Poisson rate parameters were chosen from a logarithmic distribution (rounded to the nearest integer) ranging from 10 to 1000, and the Lorentzian median and HWHM parameters were also chosen from logarithmic distributions ranging from 10 to 1000. The results of 5000 such trial draws are shown in Figure \ref{fig:etest_1D_test_histogram}, with the $E$ statistic confidence levels closely matching the expected uniform distribution. We have also made draws with a reduced range of distribution parameters, to verify that variation in the distribution parameters was not masking flaws in the confidence levels, but obtained the same uniform distribution. This suggests that our $E$ statistic confidence levels are trustworthy, and that the nonparametric nature of the statistic does indeed render it independent of the shape of the distribution.

\begin{figure*}
   \centering
   \includegraphics[width=1\textwidth]{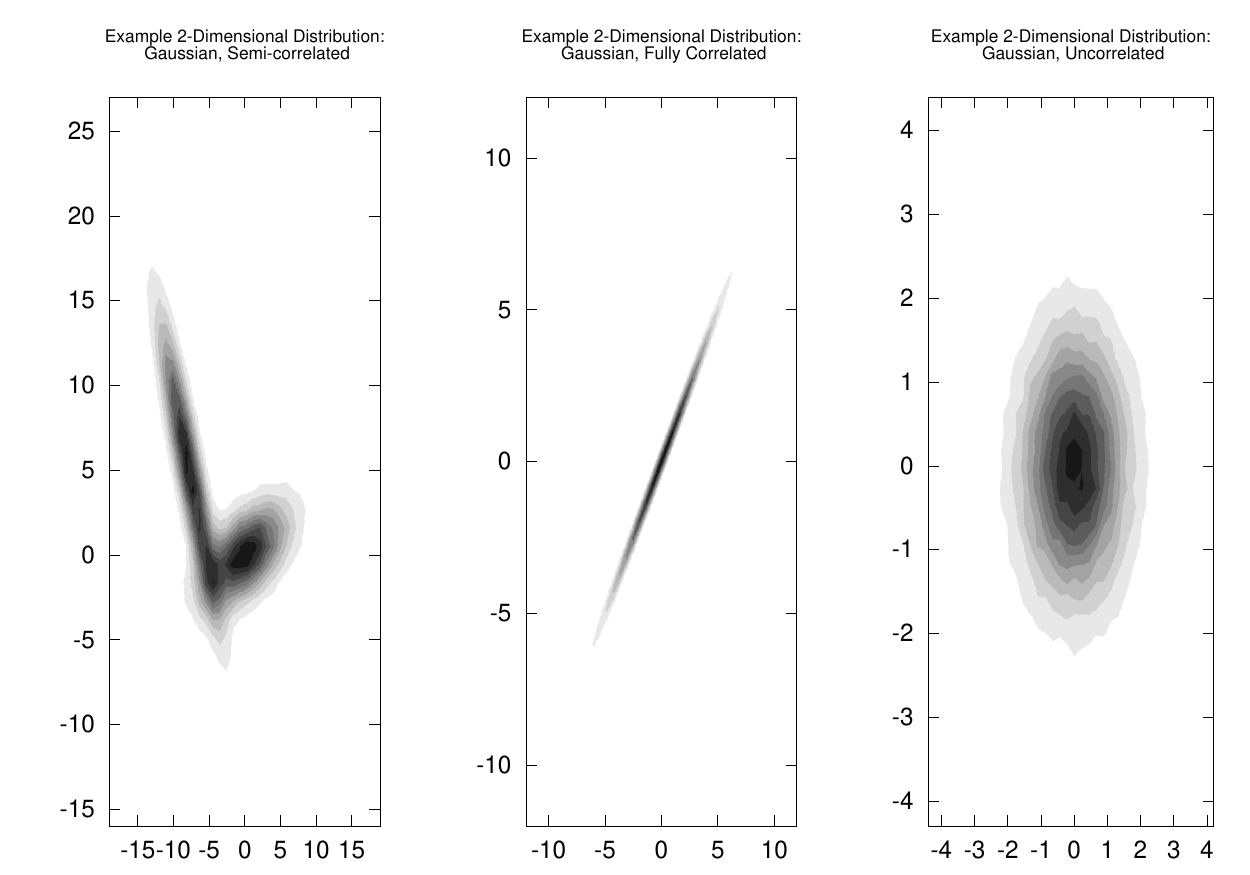}
   \caption{Contours of each of the two dimensional distributions used to test the confidence levels of the 2-D E statistic.}
   \label{fig:test_2D_distributions}
\end{figure*}

For the two dimensional tests, one significant concern \citep[see][]{Peacock83, FasanoFranceschini87} is how the confidence levels are affected by the degree of correlation in the data. We therefore use 3 test distributions. The first is an almost completely correlated (length$/$width $\sim 10^5$) Gaussian distribution along the line defined by $\theta = 45^\circ$. The second is partially correlated, and is composed of a pair of elongated two-dimensional Gaussians, partially overlapping and rotated with respect to the x and y axes. The third is uncorrelated, and consists of a single circular two dimensional Gaussian distribution. Figure \ref{fig:test_2D_distributions} shows a contour plot for each of these distributions.

For each of these distributions, we have performed 5000 comparisons of pairs of data sets drawn from the distribution. In each case, the total number of counts in the data set was Poisson distributed with $\lambda = 200$. 

\begin{figure*}
   \centering
   \includegraphics[width=1\textwidth]{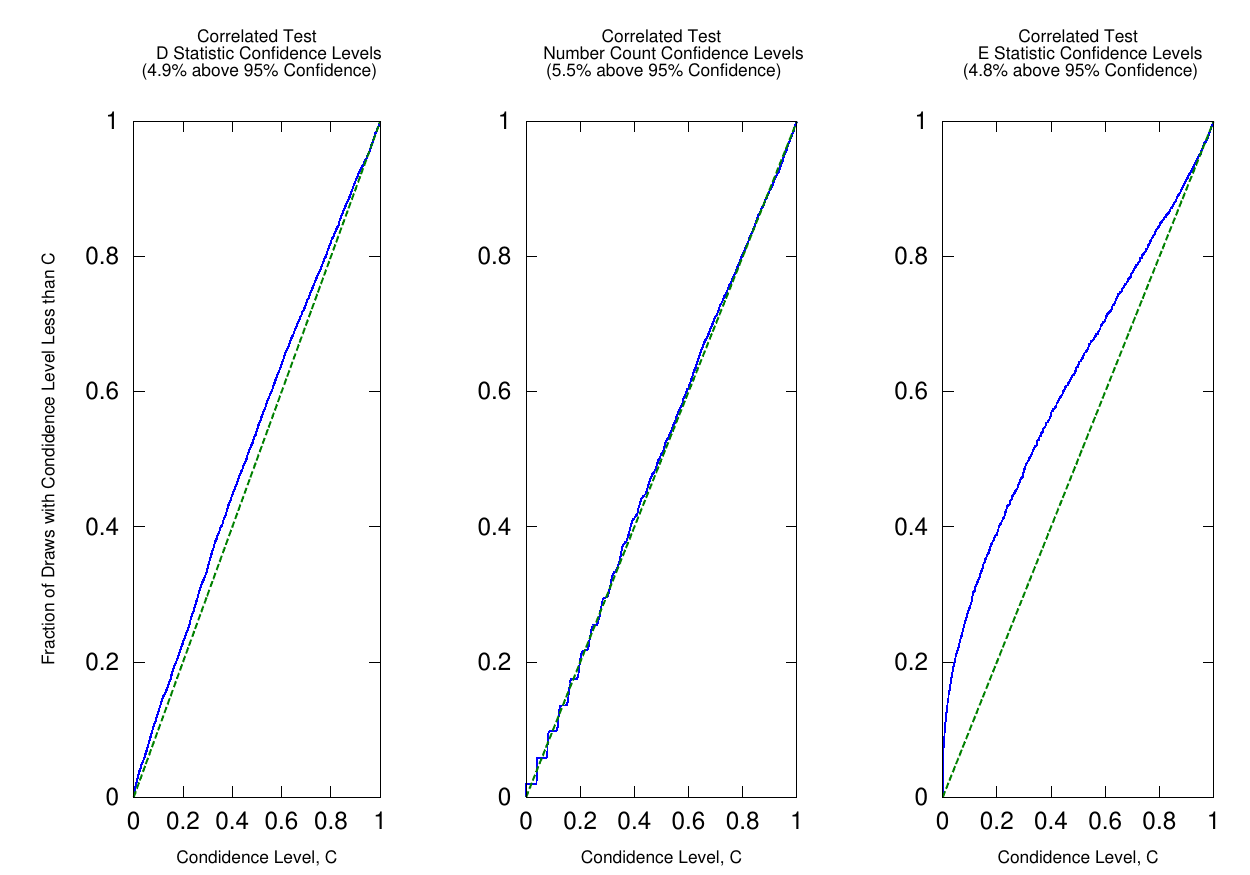}
   \caption{Cumulative histogram of confidence levels for draws given the null hypothesis with correlated distributions. Also shown is the straight line of unit slope which is the correct cumulative histogram in the limit of a large number of draws and a large number of samples in each draw. 5000 draws were made given the null hypothesis, with each data set having a Poisson total number count with parameter $\lambda = 200$. The distribution of individual data points is shown in Figure \ref{fig:test_2D_distributions}.}
   \label{fig:Etest2D_correlatedtest}
\end{figure*}

As one might expect, the $E$ test confidence levels, which were calculated using a null hypothesis with uncorrelated distributions, diverge somewhat from the ideal distribution when applied to correlated data (see Figure \ref{fig:Etest2D_correlatedtest}). However, they perform better (closer in general to the ideal distribution) than the \cite{nr} confidence levels perform for their worst case (i.e., completely {\em un}-correlated data)\footnote{Reproducing the latter distribution of confidence levels is a simple computational exercise, so we do not show it here.}. Moreover, even in the uncorrelated case they work well in the range of $95\dots 100\%$ confidence. Since only confidence levels in that range are considered statistically significant in any case, these $E$ statistic confidence levels should be suitable for use in practice. For the semi-correlated case, the accuracy is much better, and it is essentially perfect for the uncorrelated case (unsurprising since the same distribution was used to produce the confidence levels in the first place). The number count confidence levels match the ideal distribution for all cases, but this is trivially true, since it is completely insensitive to the distribution of data points, and only considers the differences between the total number counts.

We also perform the same check on each of the MBH population model results discussed in Section \ref{chap:2D_direct_comparisons} of this paper, in case the MBH parameter distributions contain features which expose problems in the confidence level estimates which are not revealed by the above tests. We find that, with 5000 comparisons made given the null hypothesis for each model and combination of model parameters, the fraction of confidence levels above 95\% is very consistently within $0.04$ and $0.06$. This indicates that our model comparison test performs correctly for the models used in this paper.

\section{Direct Application of Errors; 2-Dimensional Tests}\label{chap:2D_direct_comparisons}

In this section, we use the model comparison tests developed in section \ref{chap:modelcomp} to compare results of four models provided by Marta Volonteri. These models, which were also used in \cite{LISAPE2009}, are divided into two pairs:
\begin{itemize}
\item A pair of model results using low-mass seeds \citep[per][]{VHM03}, which differ from each other only in that one uses a standard `prolonged accretion' scenario while the other uses a chaotic accretion scenario. The two accretion scenarios give approximately identical source counts, but varying distributions in the population parameter space.
\item A second pair of results using high-mass seeds \citep[per][]{BVR}, which differ from each other in the same way as do the low-mass seed results. This high-mass seed scenario results in a significantly different source distribution than the low-mass case, with a smaller number of more massive sources.
\end{itemize}

These model results were created using the semi-analytical hierarchical SMBH merger and accretion simulation framework \citep[e.g.,][]{Coleetal00, VHM03}, specifically that which was originally developed in \cite{VHM03}. Each corresponds to a set of simulation runs using a particular combination of prescriptions for BH accretion and for the initial BH seed population. The prescriptions are listed below:

\begin{itemize}
\item In the high-mass seed prescription, seed black holes are of mass $\sim 10^5 M_\odot$ and form at redshifts $z \sim 10$ via a `quasi-star' stage.

\item In the low-mass seed prescription, seed black holes have mass $\sim 100 M_\odot$ and form at redshifts $z \sim 20$ via collapse of the first Population III stars.

\item The chaotic accretion prescription results in BHs with minimal spin due to accretion occurring in short episodes, some with co-rotating material and some with contra-rotating.

\item The prolonged, `efficient' accretion prescription results in BHs with high spin due to sustained accretion of co-aligned material.
\end{itemize}

Following \cite{LISAPE2009}, we use the following labels for the four sets of model results:
\begin{description}
\item[SC:] small seeds \citep[a l\`{a}][]{VHM03}, chaotic accretion (low spin)
\item[SE:] small seeds, efficient accretion (high spin)
\item[LC:] large seeds \citep[a l\`{a}][]{BVR}, chaotic accretion (low spin)
\item[LE:] large seeds, efficient accretion (high spin)
\end{description}
These results give an interesting range of source distributions, for purposes of examining LISA's model discriminating power. They are similar enough that it is not immediately apparent (on the basis of their event rates) that they are distinguishable, yet they differ in ways that are physically important. We use these models as test cases to investigate LISA's ability to distinguish between MBH populations. 

The simulations begin with modern-day dark matter haloes at various fiducial masses, repeatedly breaking the haloes up into progenitors at increasingly high redshift and tracking the resulting hierarchy of mergers. The process is repeated multiple times to ensure that the statistical variation in the merger tree has been adequately sampled. These high-redshift haloes are then seeded with SMBH progenitors, and the halo merger tree is then followed forward in time, applying the prescriptions for MBH merging and accretion as the simulation proceeds. Each set of model results consists of 12 files, one for each fiducial halo mass, $M_i$, which list all of the mergers that occurred during the simulation runs for that halo mass. The probability that, during some time interval $\Delta t$, a source corresponding to an present-day halo mass $M_i$ will appear as a coalescence in a realisation of the population is \citep[cf.][]{LISAPE2009}
\begin{equation}\label{psprob}
p_s(M_i,\Delta t) = 4\pi c \Delta t\Bigg[\frac{D_L(z)}{1+z}\Bigg]^2\frac{W_{PS}(M_i)}{n_t(M_i)},
\end{equation}
where $W_{PS}(M_i)$ is the `Press-Schechter Weight' of that fiducial mass, corresponding to the present-day comoving number density of sources, and $n_t(M_i)$ is the number of halo merger hierarchies that the simulation traced at fiducial halo mass $M_i$. For histograms of each model's overall distribution as a function of the population parameters (masses, luminosity distance, and spins), see the heavy black lines of Figures \ref{fig:m1detdistributions} through \ref{fig:spin2detdistributions}, below. Unless otherwise noted, mass units in this section are $M_{\odot}$, and luminosity distance is in units of Gpc.

\subsection{Direct Error Application Method}\label{sec:directerrmethod}

\begin{figure*}
   \centering
   \includegraphics[width=0.75\textwidth]{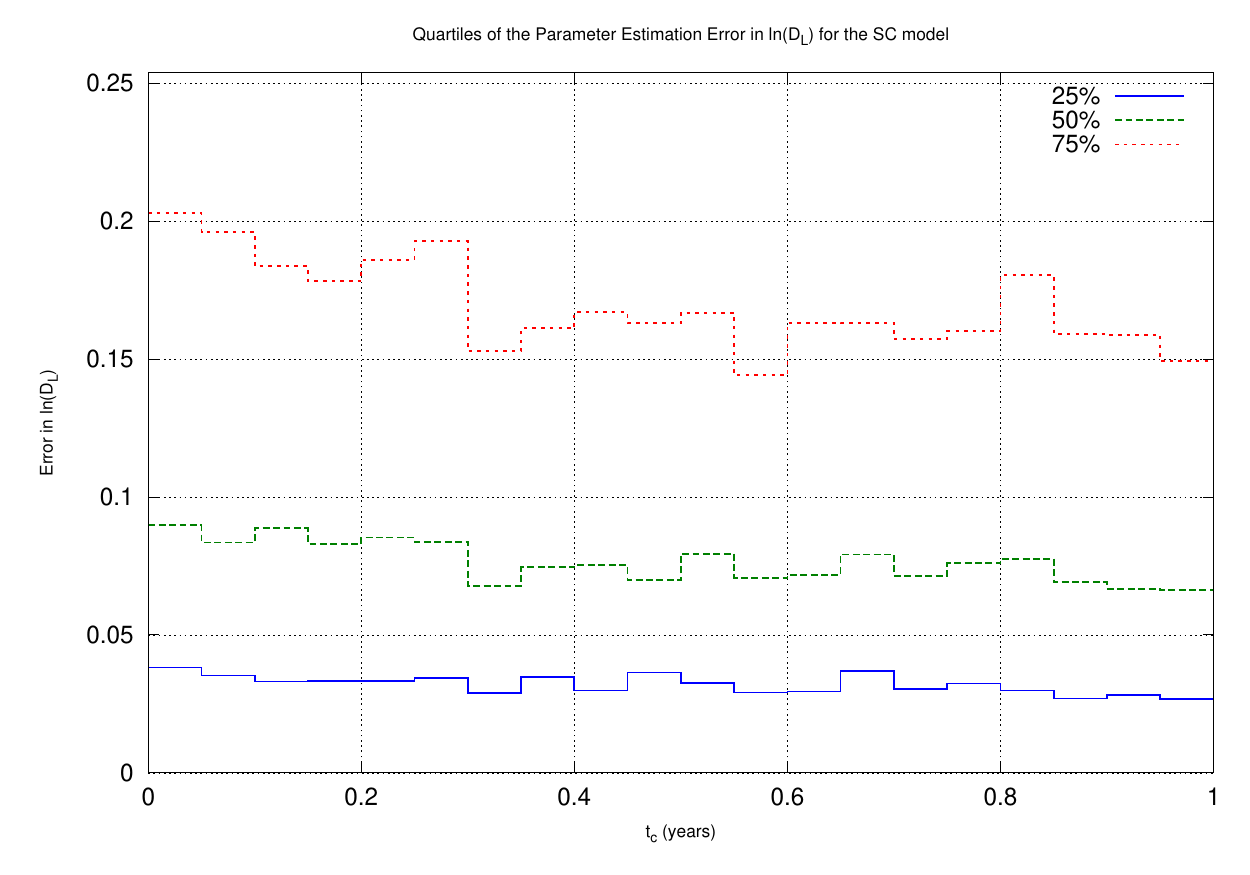}
   \caption{Quartiles of the luminosity distance estimation errors for the SC (small seeds, small spins) model. This figure illustrates the lack of dependence of the error on $t_c$, and as a result we only use error distributions produced with a one year range of $t_c$, even for longer observation windows.}
   \label{fig:tc_dlerror}
\end{figure*}

Due to the much larger number of bins required to cover a 5-dimensional parameter space and the far slower execution time of the code used to calculate parameter estimation errors of spinning BH binaries, the error kernel method used in \cite{plowetal2009I} is not practical for the distributions studied here. Rather than producing an error kernel in the 5-dimensional population parameter space (masses, luminosity distance, and spins) therefore, we compute and apply the parameter estimation errors with a simpler, more direct method. It proceeds as follows:

To each source in the model result files mentioned above, we assign a randomly generated set of sample parameters (sky positions, orientations, spin orientations). In practice, we found that the very small spins of the SC and LC distributions caused unrealistically high spin estimation errors due to the Fisher matrix becoming singular for zero spin. To alleviate this issue, we have altered the spin distributions of these models so that spins which were below 0.01 instead lie between 0.01 and 0.1. For the range of the time to coalescence, $t_c$, we only employ a 1 year range for all draws from the models, rather than producing new draws with a larger range of $t_c$ when we draw sources for the 3 and 5 year observations discussed below. The dependence of the errors on $t_c$ appears to be minor, as illustrated by Figure \ref{fig:tc_dlerror}.

We then calculate the SNR and parameter estimation errors of each source, using both the Fisher Information matrix code mentioned in Section \ref{sec:BBHparams} and the LISA Calculator used in \cite{plowetal2009I}. The LISA calculator \citep{LC} is first used to obtain a rough estimate of the source's SNR, and the more advanced (but much slower) spinning BH code \citep{LISAPE2009} is used to calculate the errors of each source which has a LISA calculator SNR $\geq 5$. The LISA calculator used a galactic foreground confusion noise estimate from \cite{TRC06}, while the spinning BHB code used the instrument and confusion noise model listed in Section 4.2 of \cite{LISAPE2009}.

We then store the source parameters, SNRs, and parameter estimation uncertainties (i.e., the a priori standard deviations in the estimated parameters for that set of source parameters) to disk. Thus, for each of the model result files mentioned above, we have a corresponding file that lists a full set of binary parameters (including randomly generated sample parameters), binary SNRs, and parameter estimation errors for every source in the model result file. For each of the low-mass seed models (SC and SE), there are $\approx 150000$ sources, while the high mass seed models have $\approx 40000$ sources.

Next, we generate a model predicted realisation of the set of sources occurring during some observation window $\Delta t_{obs}$. We run through each source occurring in the model's result files, performing a Bernoulli trial with probability of success given by equation \ref{psprob} to determine whether that source will occur in the realisation. For each of those sources, we look up its previously calculated parameter estimation errors and draw from the corresponding Gaussian distributions\footnote{If resulting parameters fall outside of the possible ranges of these parameters (e.g., the spin parameters lie between 0 and 1, by definition), the parameters are wrapped around so that they remain within the range. We also add a small $10^{-3}$ Gaussian variation to the source parameters, to avoid issues with the model comparison caused by sources grouped at a single parameter value.}. This produces a realisation of the model predicted best-fit parameters, as estimated by LISA, for sources occurring within the observation window.

We then calculate these sets of estimated parameters for a pair of population models, and compare them using the statistical tests discussed in Section \ref{chap:modelcomp} to see whether or not there is a statistically significant difference between them. We repeat the experiment many times to control for statistical variation in the comparison results, and we investigate the distinguishability of each combination of models and choices of population parameters. We also consider how the increased number of sources afforded by a longer (i.e., 3 year, as opposed to 1 year) observation window affects the distinguishability of the models, and compare the LISA best-fit parameter realisations to realisations with no parameter estimation errors, to see if these errors have a significant effect on model distinguishability. The results of this work are presented in the next section.

\subsection{The Parameter Estimation Errors \& Estimated Parameter Distributions}\label{sec:parmerrors}
In this section, we list the distribution of parameter estimation errors for each model. The extent to which these errors affect the estimated parameter distributions will depend on whether or not the errors are larger or smaller than the feature sizes of the distribution to which they are applied. When the source distribution has features smaller than the errors, they will have a significant effect on the parameter distribution. In that case, the estimated parameter distribution can be confused with some other source distribution when they would otherwise (i.e., with better parameter determination) be distinguishable.

\begin{figure*}
   \centering
   \includegraphics[width=0.75\textwidth]{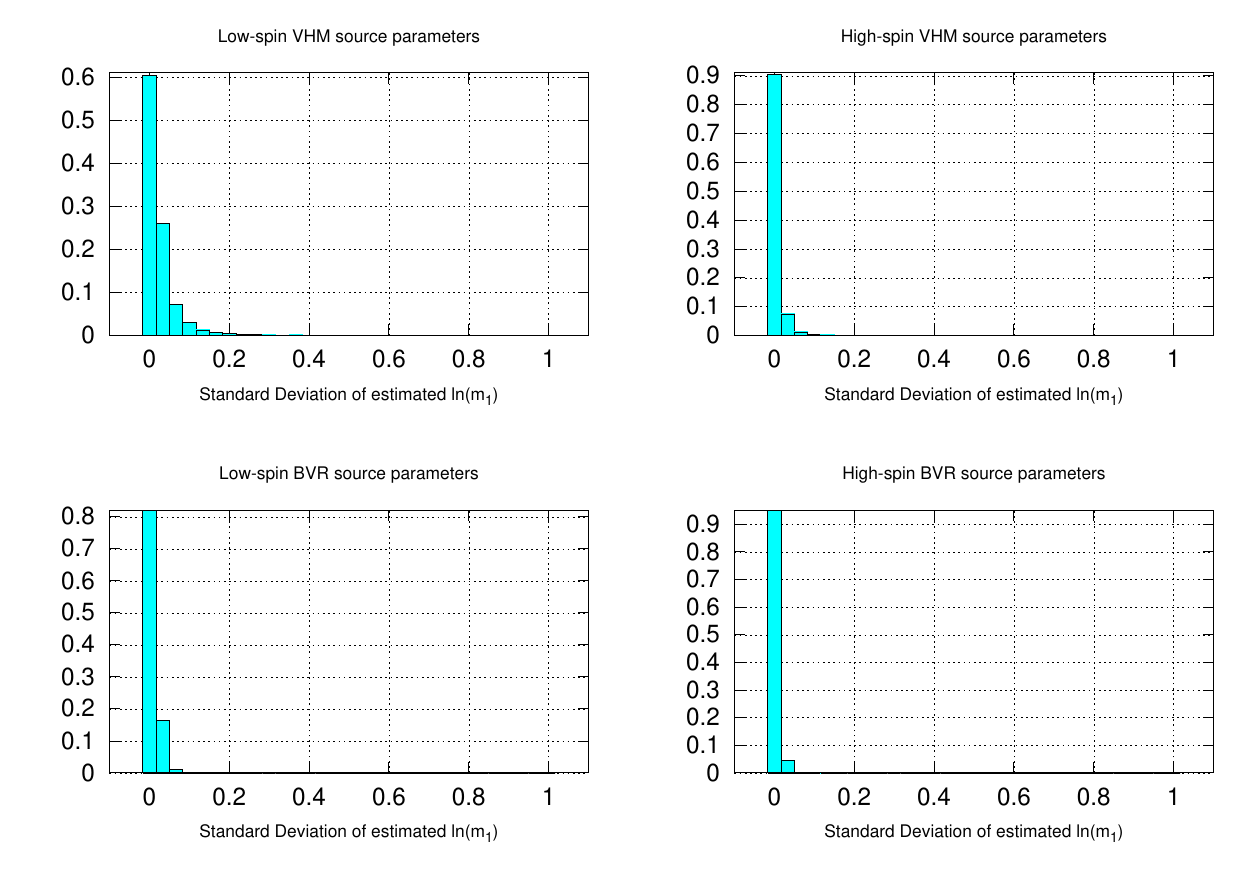}
   \caption{The parameter estimation error in $\ln{m_1}$ (the redshifted mass of the larger black hole) for binaries detected by LISA.}
   \label{fig:m1errdistributions}
\end{figure*}

\begin{figure*}
   \centering
   \includegraphics[width=0.75\textwidth]{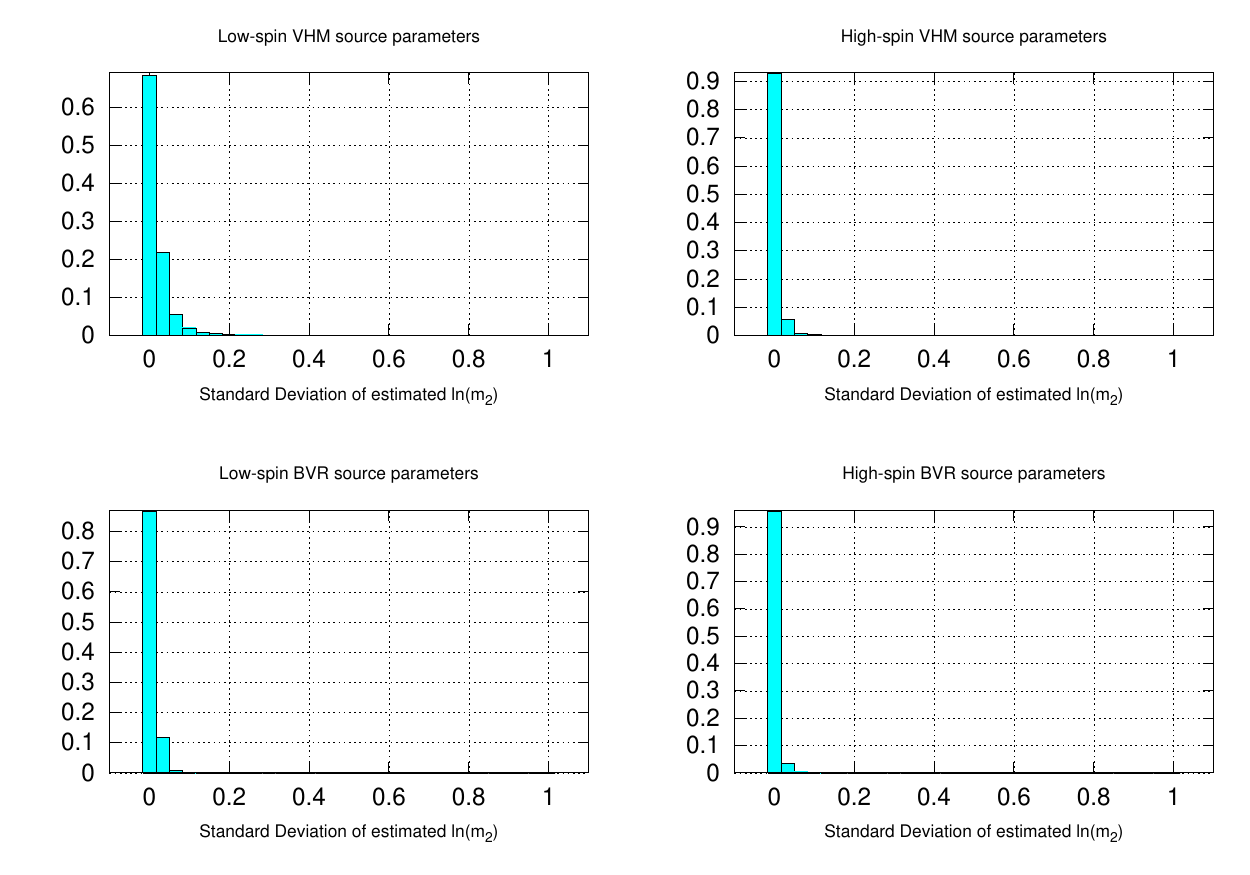}
   \caption{The parameter estimation error in $\ln{m_2}$ (the redshifted mass of the smaller black hole) for binaries detected by LISA.}
   \label{fig:m2errdistributions}
\end{figure*}

\begin{figure*}
   \centering
   \includegraphics[width=0.75\textwidth]{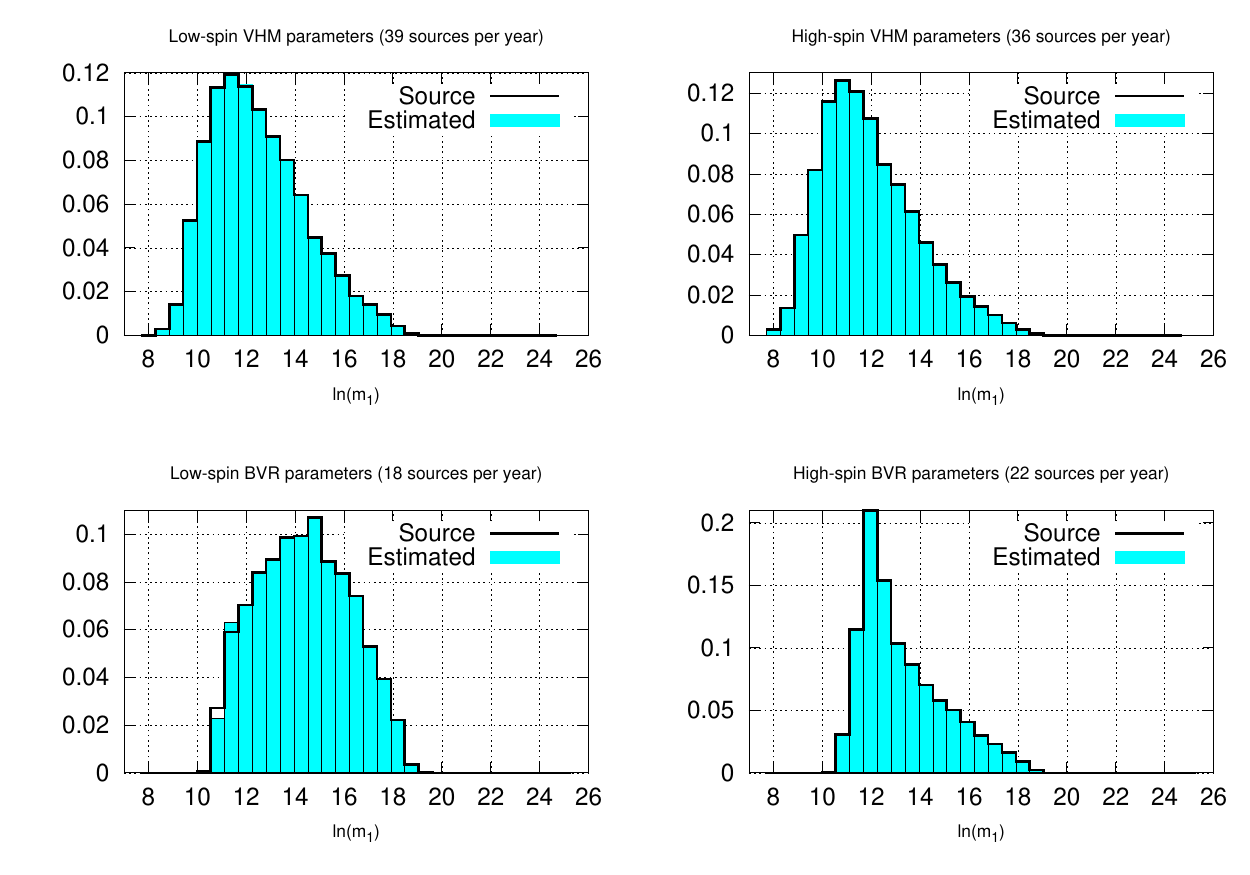}
   \caption{The estimated parameter distribution of $m_1$ (the redshifted mass of the larger black hole) for binaries detected by LISA. Also shows the corresponding source parameter distribution (an SNR=10 cutoff has been applied to both distributions).}
   \label{fig:m1detdistributions}
\end{figure*}

\begin{figure*}
   \centering
   \includegraphics[width=0.75\textwidth]{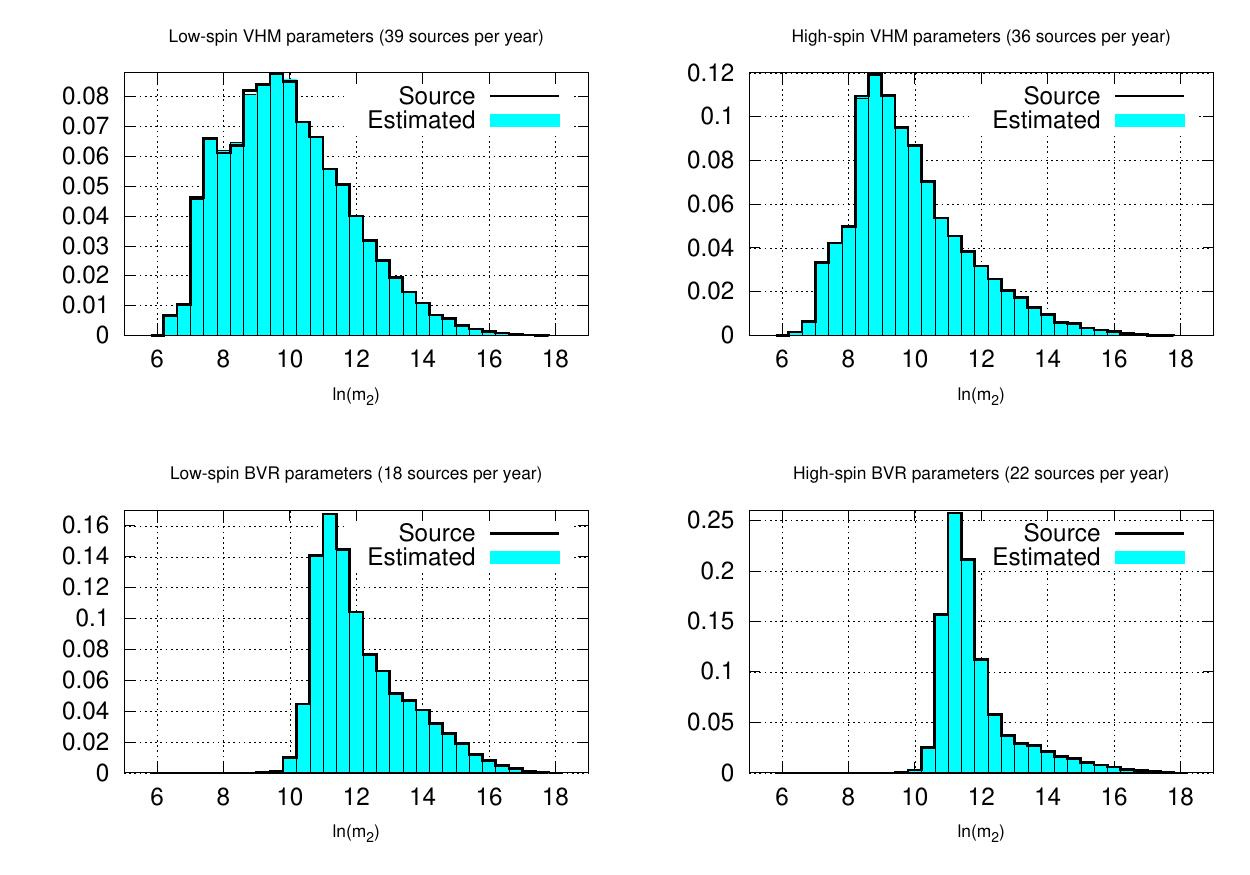}
   \caption{The source parameter distribution of $m_2$ (the redshifted mass of the smaller black hole) for binaries detected by LISA.  Also shows the corresponding source parameter distribution (an SNR=10 cutoff has been applied to both distributions).}
   \label{fig:m2detdistributions}
\end{figure*}

When compared to the characteristic feature sizes of the model distributions (see Figures \ref{fig:m1detdistributions} and \ref{fig:m2detdistributions}), the parameter estimation errors in the masses (Figures \ref{fig:m1errdistributions} and \ref{fig:m2errdistributions}) are quite small for all of the models involved. By that measure, we do not expect that the parameter estimation errors in the (redshifted) masses to have a significant effect on LISA's ability to distinguish between the models. 

\begin{figure*}
   \centering
   \includegraphics[width=0.75\textwidth]{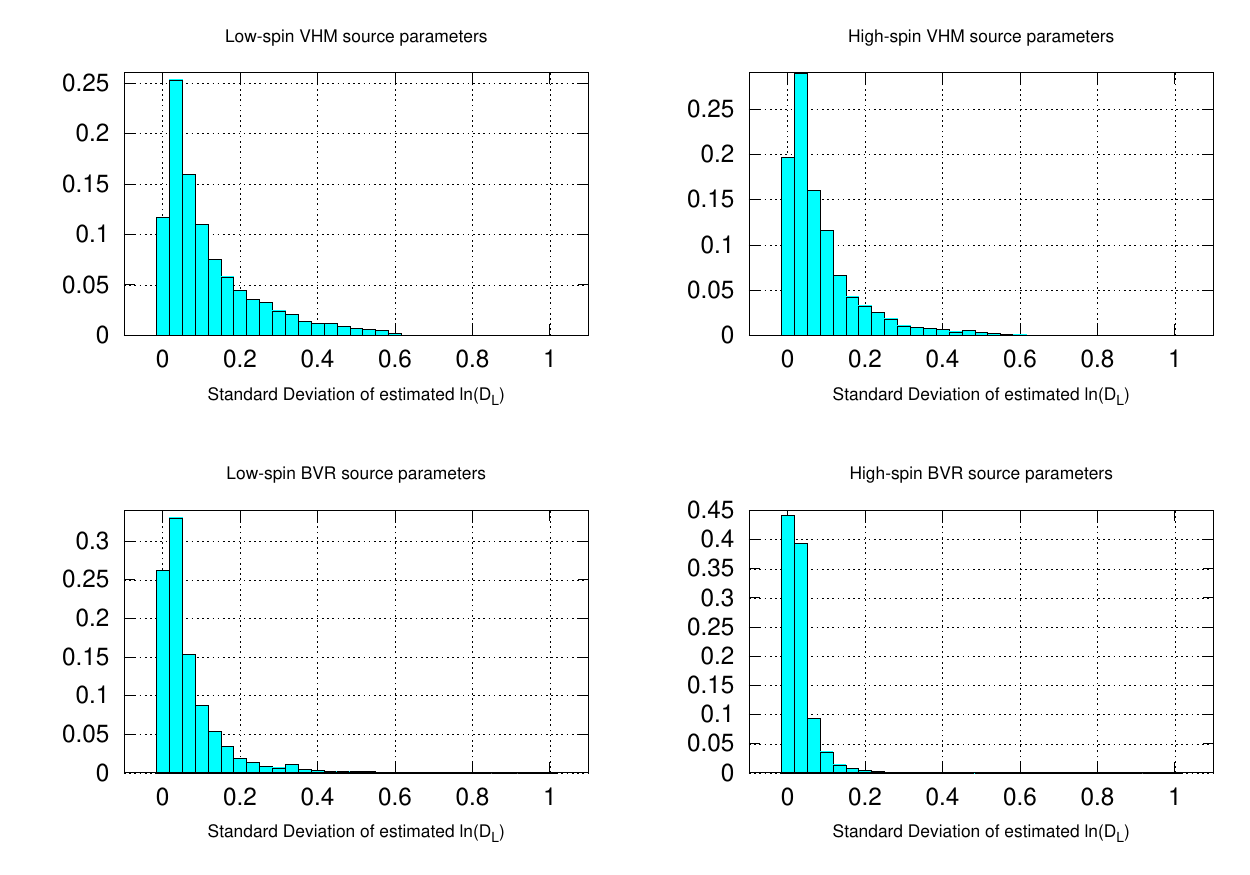}
   \caption{The parameter estimation error in $\ln{D_L}$ for binaries detected by LISA.}
   \label{fig:DLerrdistributions}
\end{figure*}

\begin{figure*}
   \centering
   \includegraphics[width=0.75\textwidth]{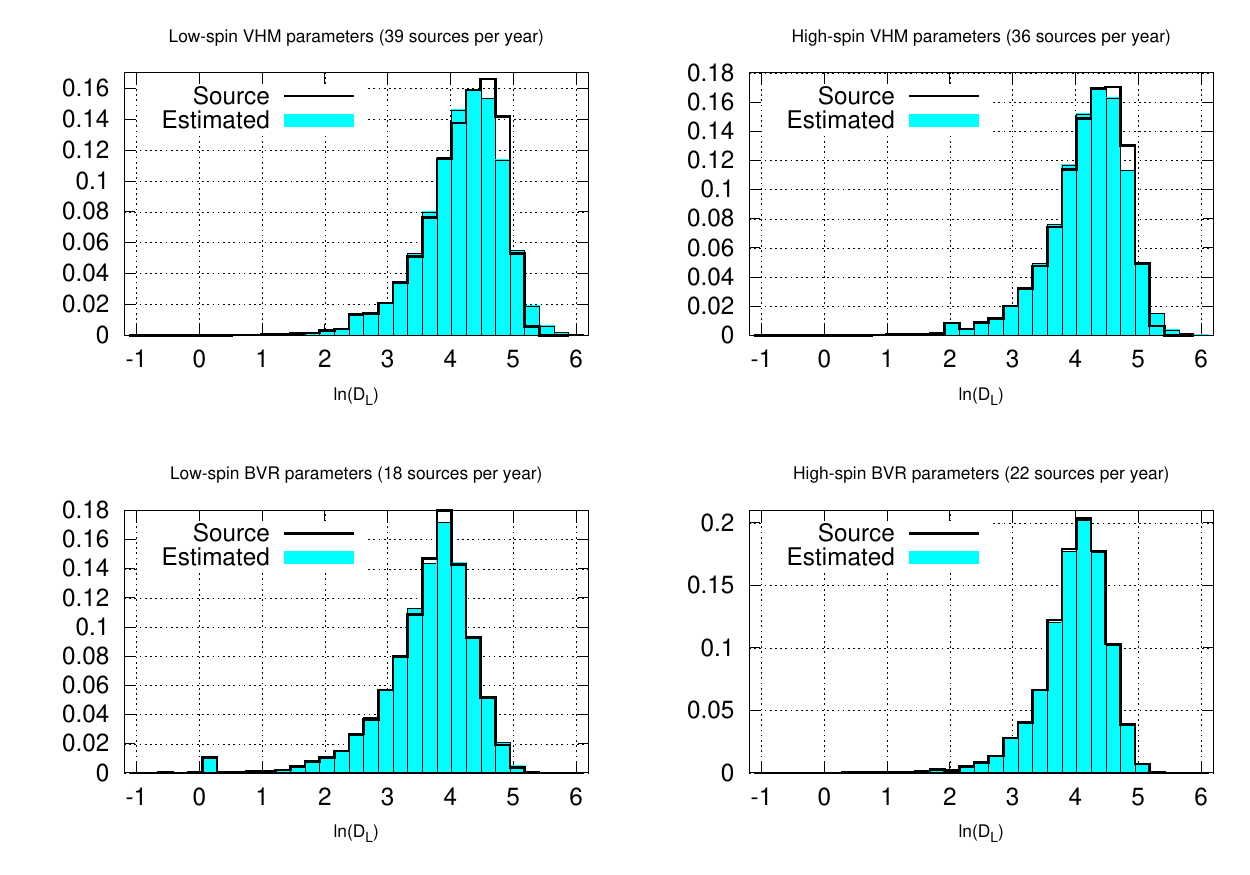}
   \caption{The source parameter distribution of $D_L$ for binaries detected by LISA.  Also shows the corresponding source parameter distribution (an SNR=10 cutoff has been applied to both distributions).}
   \label{fig:DLdetdistributions}
\end{figure*}

The luminosity distance errors (Figure \ref{fig:DLerrdistributions}), on the other hand, are somewhat large compared to the characteristic feature sizes of the models (see Figure \ref{fig:DLdetdistributions}), so they may have an effect on the model distinguishability. 

\begin{figure*}
   \centering
   \includegraphics[width=.8\textwidth]{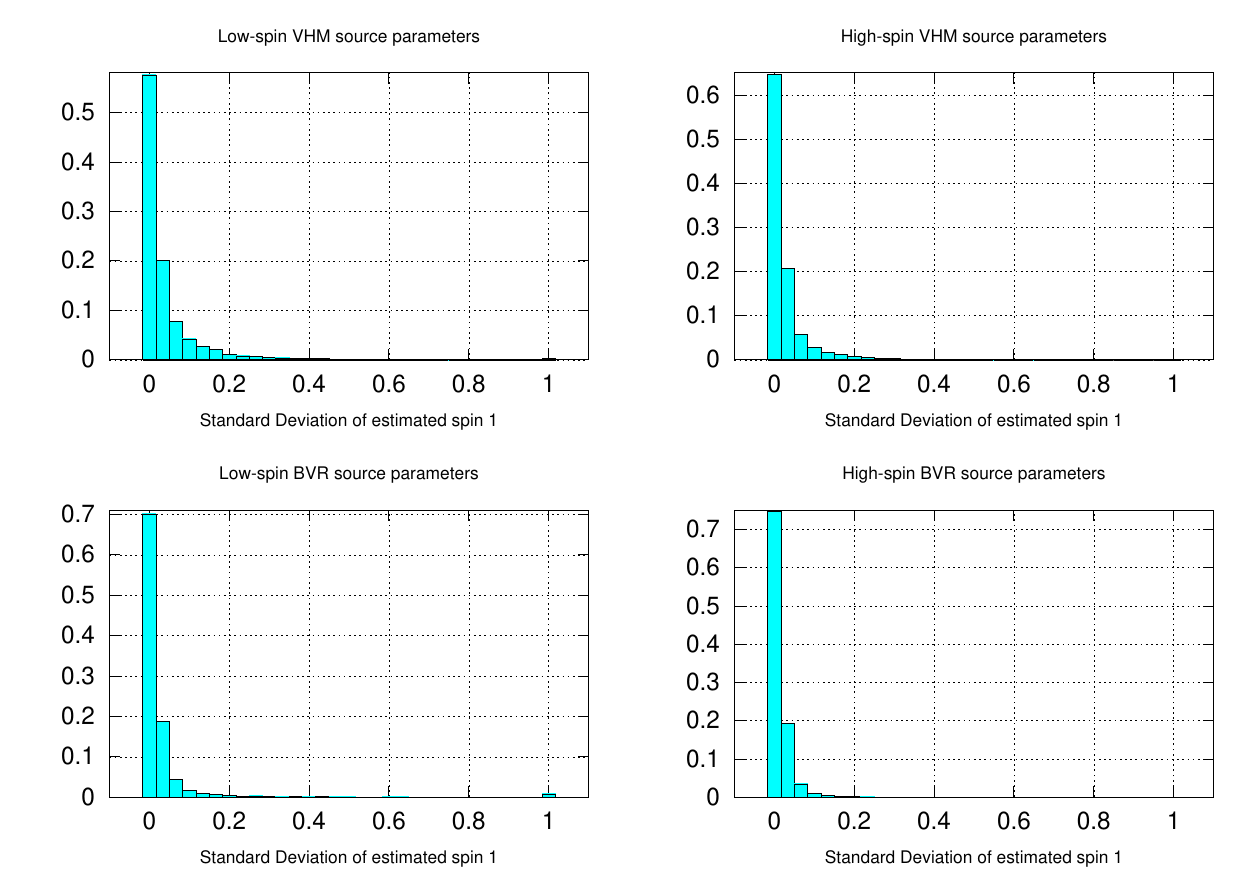}
   \caption{The parameter estimation error in the spin of the larger black hole for binaries detected by LISA.}
   \label{fig:spin1errdistributions}
\end{figure*}

\begin{figure*}
   \centering
   \includegraphics[width=.8\textwidth]{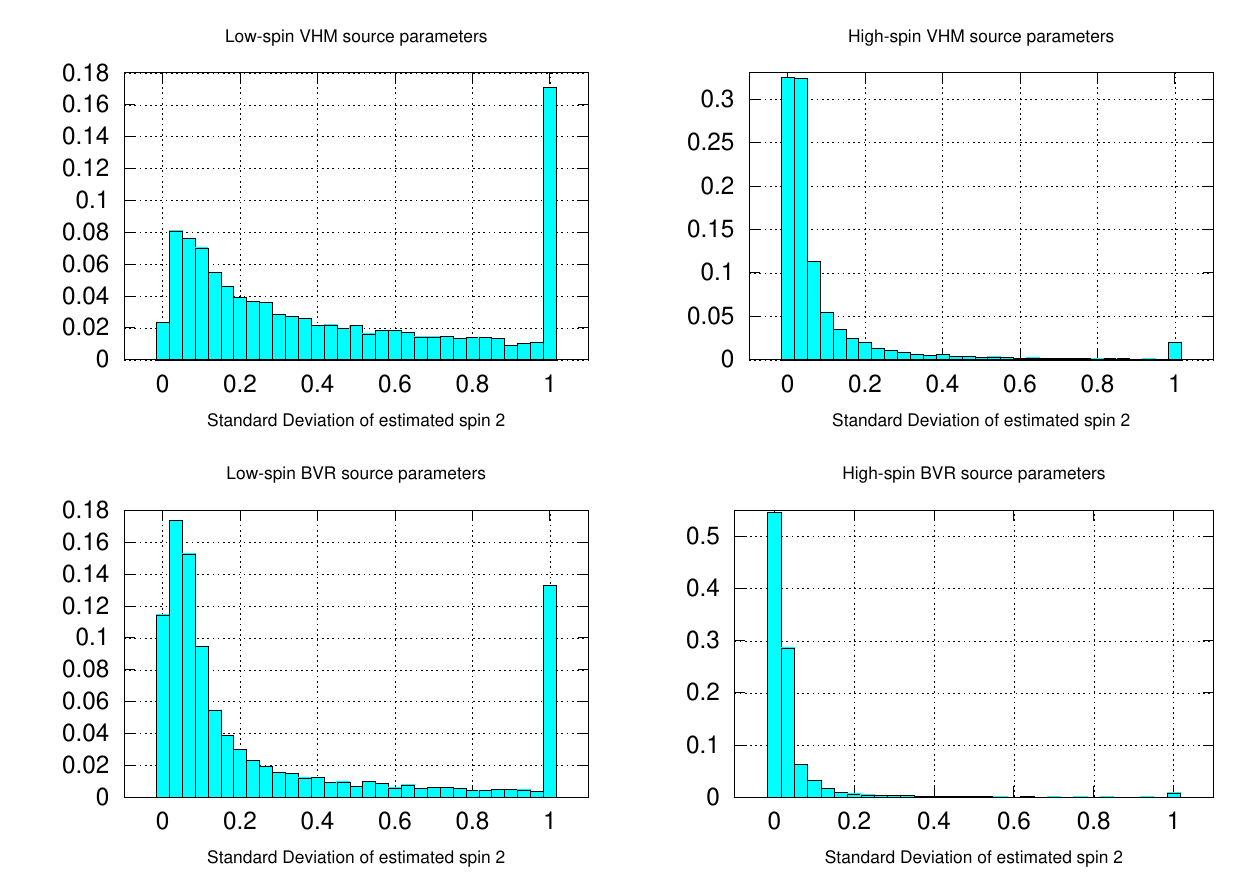}
   \caption{The parameter estimation error in the spin of the smaller black hole for binaries detected by LISA.}
   \label{fig:spin2errdistributions}
\end{figure*}

\begin{figure*}
   \centering
   \includegraphics[width=0.75\textwidth]{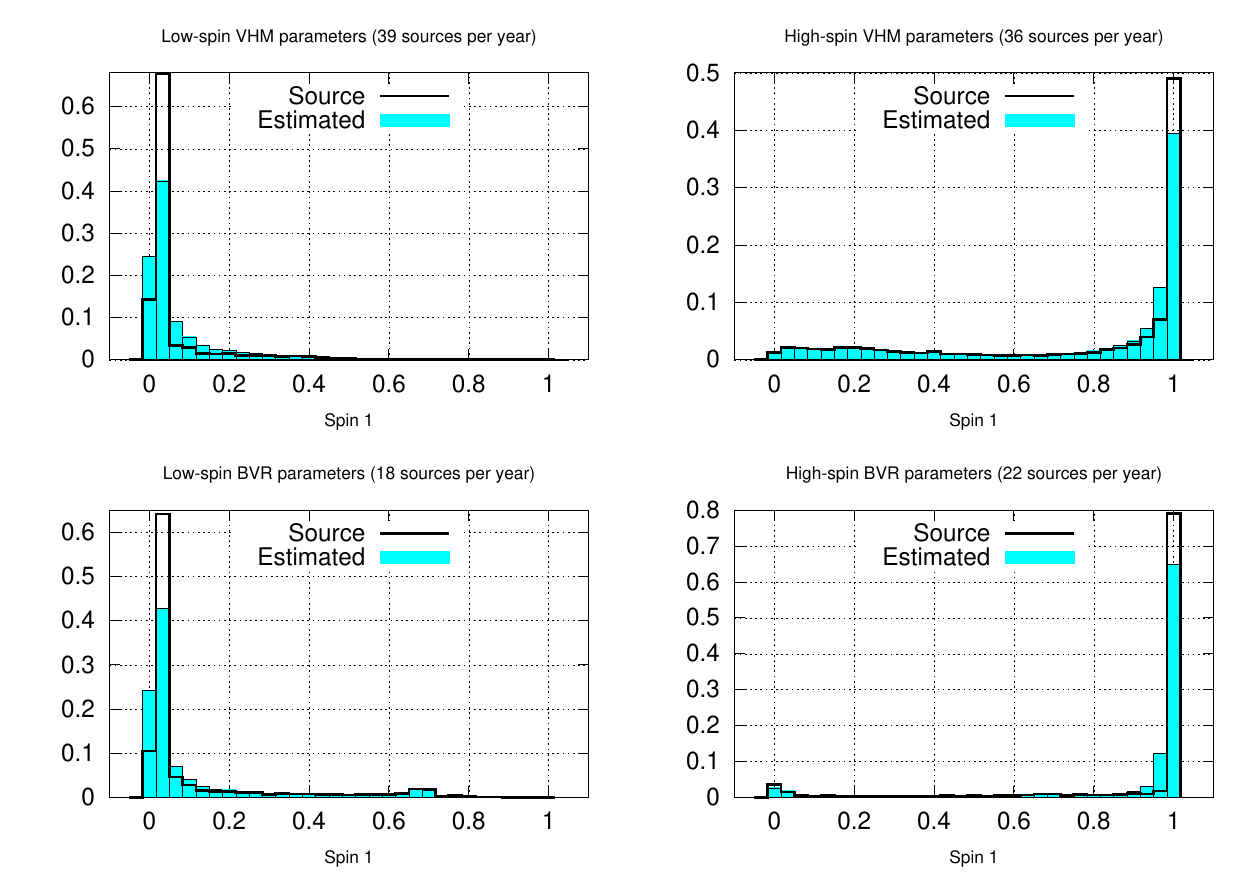}
   \caption{The source parameter distribution of the spin of the larger black hole for binaries detected by LISA.  Also shows the corresponding source parameter distribution (an SNR=10 cutoff has been applied to both distributions).}
   \label{fig:spin1detdistributions}
\end{figure*}

\begin{figure*}
   \centering
   \includegraphics[width=0.75\textwidth]{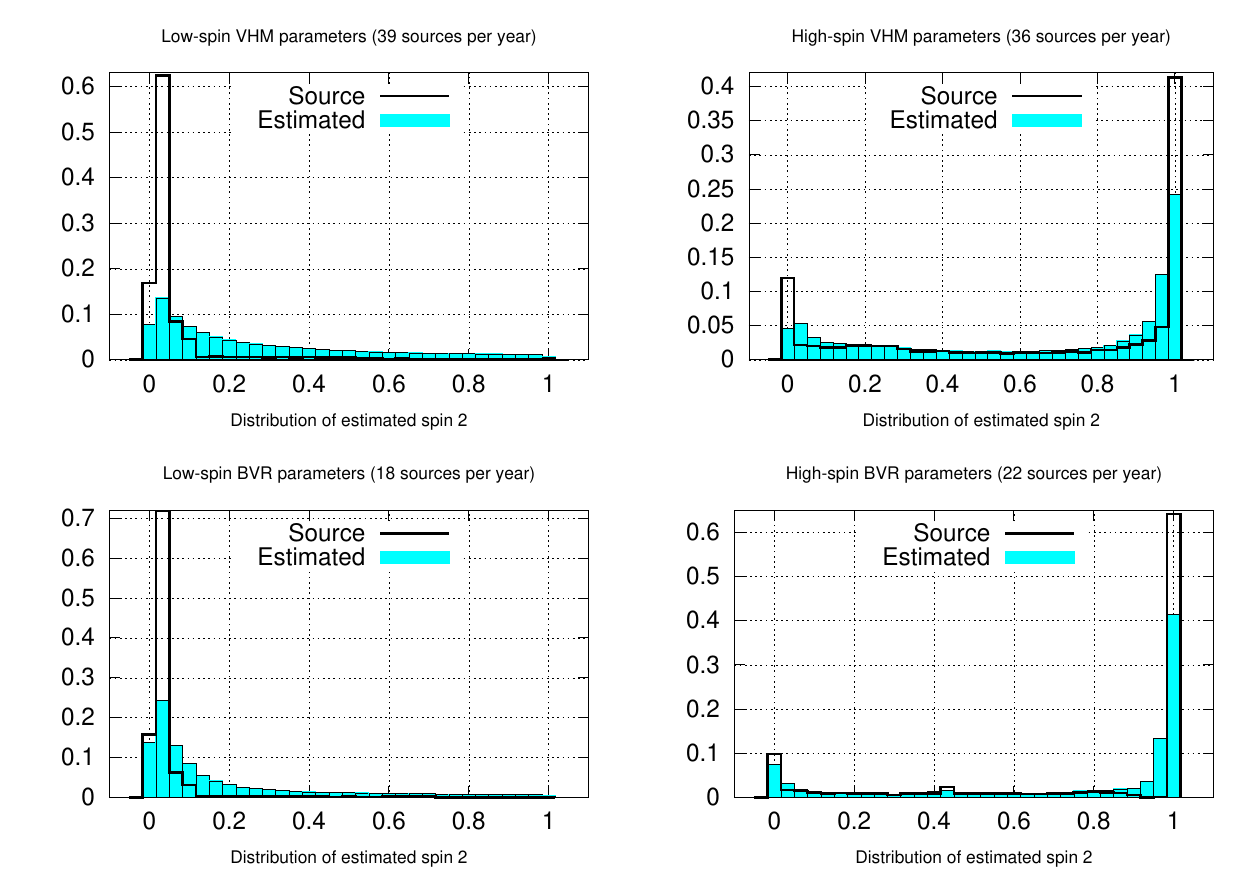}
   \caption{The source parameter distribution of the spin of the smaller black hole for binaries detected by LISA.  Also shows the corresponding source parameter distribution (an SNR=10 cutoff has been applied to both distributions).}
   \label{fig:spin2detdistributions}
\end{figure*}

The spin errors are also both large {\em compared to} the model feature sizes, but it is less certain how meaningful the result is in this case. In the case of the high-spin models, the spin distributions of the sources (see Figures \ref{fig:spin1detdistributions} and \ref{fig:spin2detdistributions}) are extremely tightly clustered (most of the spins parameters of the larger BH are equal to 0.998\footnote{Based on an estimated maximum from \cite{Thorne74}.}), so that even though the spin errors (Figures \ref{fig:spin1errdistributions} and \ref{fig:spin2errdistributions}) are quite small, the parameter distribution cannot be constrained as tightly as the model's predicted distribution. For the low-spin parameters, the parameter estimation errors are unrealistically high, due to the linear assumptions of the Fisher information matrix approach, and the actual parameters may be better constrained due to higher order corrections to the likelihood surface. The large size of the error in spin 2 (i.e., $\chi_2$) is not surprising when we remember that the spin angular momentum of a black hole is given by $\chi m^2$ (mentioned in Section \ref{sec:BBHparams}) and typical mass ratios are $\sim 0.1$, so that spin angular momentum of the smaller BH is only a small fraction of the angular momentum of the system (recall that $\chi \le 1$).

The sizes of these errors are summarised in Table \ref{tab:errorquartiles}, which shows the third quartile parameter estimation error for each combination of model and parameter. For instance, the entry in the `SC' column and the `$\ln{D_L}$' row shows that $3/4$, or $75\%$ of the parameter estimation errors in $\ln{D_L}$ for the small seed, chaotic accretion model are distributed with a standard deviation of $0.168$ or less

\begin{table}
\caption{Third quartiles ($75$th percentiles) of the standard deviation of estimated parameters for each of the four models considered.}\label{tab:errorquartiles}
\centering
\begin{tabular}{c|c|c|c|c}
                &         SC &         SE &         LC &         LE \\
\hline\hline
     $\ln{m_1}$ &      0.029 &      0.006 &      0.014 &      0.006 \\
\hline
     $\ln{m_2}$ &      0.022 &      0.005 &      0.011 &      0.005 \\
\hline
     $\ln{D_L}$ &      0.168 &      0.112 &      0.084 &      0.036 \\
\hline
         Spin 1 &      0.042 &      0.027 &      0.021 &      0.017 \\
\hline
         Spin 2 &      0.766 &      0.079 &      0.397 &      0.033 \\
\hline
\end{tabular}
\end{table}

Figures \ref{fig:m1detdistributions} through \ref{fig:spin2detdistributions} show the distribution of best-fit parameters which are obtained when random estimated parameters are drawn from the distribution of each source's parameter estimation errors, as described in Section \ref{sec:directerrmethod}. Each plot also shows, for comparison, the corresponding source parameter distributions. 

As can be seen from the figures, the parameter estimation errors have no significant effect on the distribution of mass variables. The $D_L$ estimation errors, on the other hand, do have some effect on the distribution of the luminosity distance in the case of the low spin VHM model (the `SE' model), and a smaller effect on the higher spin VHM model. The luminosity distance errors do not appear to have a significant effect on the parameters of the two BVR models (LE and LC), which is unsurprising given the larger masses and closer distances of those sources. For the low-spin models (SC and LC), the parameter estimation errors lead to a very significant difference in the spin distribution of the smaller BH, and a somewhat significant difference in the spin of the larger BH. For the high-spin models (SE and LE), the effects of the spin determination errors are less significant, although they still have an effect on the parameter distribution due to the model's tight clustering around a spin of 0.998 (also mentioned above).

\subsection{Model Comparison Results; 1 and 2 Dimensions}

This section reports the results of comparisons between each of the models using the tests developed in Section \ref{chap:modelcomp}. For each pair of models and model parameters, we have made 5000 comparisons as described in Section \ref{sec:directerrmethod}. We quantify the degree of distinguishability of the models by the fraction of comparisons which resulted in a 95\% or higher confidence level that the models were not drawn from the same distribution. These fractions can be understood as follows:
\begin{itemize}
\item If $9/10$ of comparisons resulted in a confidence level above $95\%$, the models can be considered `consistently distinguishable'.
\item If $3/4$ of comparisons resulted in a confidence level above $95\%$, the models can be considered `often distinguishable'.
\item If $1/2$ of comparisons resulted in a confidence level above $95\%$, the models can be considered `sometimes distinguishable'.
\end{itemize}
Table \ref{tab:comptable_1yr} shows the resulting fractions for comparison of the distributions of detected parameters and one year of observation. All of the models can be consistently distinguished based on some combination of one or two of their parameter distributions:
\begin{itemize} 
\item The SE and SC (High and Low spin VHM) models can be consistently distinguished based on the distribution of spin 1, unsurprisingly (see Figure \ref{fig:spin1detdistributions}). 
\item The LC and SC (Low spin VHM and BVR) models can be consistently distinguished based on the distributions of $\ln{m_1}$ (Figure \ref{fig:m1detdistributions}) and $\ln{m_2}$ (Figure \ref{fig:m2detdistributions}).
\item The LC and SE (Low spin BVR and high spin VHM) models can be consistently distinguished based on almost any combination of parameters. The lone exception is the distribution of $\ln{D_L}$ (Figure \ref{fig:DLdetdistributions}), but even in that case the models can often be distinguished.
\item The LE and SC (High spin BVR and low spin VHM) models can be consistently distinguished based on the $\ln{m_2}$ or spin distributions (Figures \ref{fig:m2detdistributions}, \ref{fig:spin1detdistributions}, and \ref{fig:spin2detdistributions}).
\item The LE and SE (High spin BVR and VHM) models can be consistently distinguished based on their distribution of $\ln{m_2}$ (Figure \ref{fig:m2detdistributions}). The $\ln{m_1}$ distributions, on the other hand are only sometimes distinguishable.
\item The LE and LC (High and low spin BVR) models, like the SE and SC models, can be consistently distinguished based on the distribution of spin 1.
\end{itemize}
Naturally, these numbers improve when a longer observation window affords more sources to make the comparison. Table \ref{tab:comptable_3yr} shows comparison results when sources are chosen from equation \ref{psprob} according to a 3 year observation window. With the 3 years of sources, the only model parameter distributions that cannot be reliably distinguished are the SE vs. SC distributions and the LE vs. LC distributions, as functions of the masses and $D_L$. The SE and SC distributions of masses and $D_L$, in particular, are not generally distinguishable. The distributions of $\ln{D_L}$ and $\ln{m_1}$ for the LE and LC models, on the other hand, are often distinguishable.

\begin{table*}\footnotesize
\caption{E statistic model comparison results for 1 year of observation, showing the fraction of confidence levels which were above 95\%. 
%To make the table easier to read, comparisons where $1/2$ of the confidence levels were above 95\% are marked in light grey, comparisons where $3/4$ were above 95\% are marked in medium grey, and those where $9/10$ were above 95\% are marked in dark grey.
}\label{tab:comptable_1yr}
\centering
\begin{tabular}[h]{c|c|c|c|c|c|c}

                         &   SE vs.   SC  &   LC vs.   SC  &   LC vs.   SE  &   LE vs.   SC  &   LE vs.   SE  &   LE vs.   LC  \\
\hline\hline
              $\ln{m_1}$ &         0.0776 & \cclrdg  0.918 & \cclrdg 0.9274 & \cclrlg  0.566 & \cclrlg 0.5992 &          0.163 \\
\hline
$\ln{m_2}$ \& $\ln{m_1}$ &          0.086 & \cclrdg  0.989 & \cclrdg 0.9926 & \cclrdg 0.9716 & \cclrdg 0.9878 &         0.1982 \\
\hline
              $\ln{m_2}$ &         0.0618 & \cclrdg  0.989 & \cclrdg  0.991 & \cclrdg 0.9842 & \cclrdg 0.9948 &         0.1148 \\
\hline
$\ln{D_L}$ \& $\ln{m_1}$ &         0.0894 & \cclrdg 0.9504 & \cclrdg 0.9532 & \cclrlg 0.5296 & \cclrlg 0.5502 &         0.2964 \\
\hline
$\ln{D_L}$ \& $\ln{m_2}$ &         0.0716 & \cclrdg 0.9936 & \cclrdg 0.9946 & \cclrdg 0.9678 & \cclrdg 0.9876 &         0.2376 \\
\hline
              $\ln{D_L}$ &         0.0562 & \cclrgr 0.8956 & \cclrgr 0.8422 & \cclrlg 0.5336 &         0.4366 &         0.1834 \\
\hline
    Spin 1 \& $\ln{m_1}$ & \cclrdg      1 & \cclrdg 0.9154 & \cclrdg      1 & \cclrdg      1 & \cclrlg 0.6742 & \cclrdg 0.9854 \\
\hline
    Spin 1 \& $\ln{m_2}$ & \cclrdg      1 & \cclrdg 0.9774 & \cclrdg      1 & \cclrdg      1 & \cclrdg 0.9964 & \cclrdg 0.9758 \\
\hline
    Spin 1 \& $\ln{D_L}$ & \cclrdg      1 & \cclrgr 0.8912 & \cclrdg 0.9992 & \cclrdg      1 & \cclrlg 0.5204 & \cclrdg 0.9674 \\
\hline
                  Spin 1 & \cclrdg      1 & \cclrgr  0.784 & \cclrdg 0.9996 & \cclrdg      1 & \cclrlg 0.5086 & \cclrdg 0.9808 \\
\hline
    Spin 2 \& $\ln{m_1}$ & \cclrgr 0.7958 & \cclrdg 0.9574 & \cclrdg 0.9982 & \cclrdg 0.9792 & \cclrlg 0.6928 & \cclrgr  0.892 \\
\hline
    Spin 2 \& $\ln{m_2}$ & \cclrgr 0.7882 & \cclrdg 0.9898 & \cclrdg      1 & \cclrdg 0.9992 & \cclrdg 0.9944 & \cclrgr 0.8186 \\
\hline
    Spin 2 \& $\ln{D_L}$ & \cclrgr 0.7638 & \cclrdg 0.9086 & \cclrdg 0.9696 & \cclrdg 0.9622 & \cclrlg 0.5136 & \cclrgr 0.7682 \\
\hline
    Spin 2 \&     Spin 1 & \cclrdg      1 & \cclrgr  0.836 & \cclrdg 0.9996 & \cclrdg      1 & \cclrlg   0.59 & \cclrdg 0.9918 \\
\hline
                  Spin 2 & \cclrgr 0.7674 & \cclrgr 0.8328 & \cclrdg 0.9646 & \cclrdg  0.983 &          0.494 & \cclrgr  0.786 \\
\hline
\end{tabular}
\end{table*}

\begin{table}\footnotesize
\caption{E statistic model comparison results for 3 years of observation, showing the fraction of confidence levels which were above 95\%. 
%To make the table easier to read, comparisons where $1/2$ of the confidence levels were above 95\% are marked in light grey, comparisons where $3/4$ were above 95\% are marked in medium grey, and those where $9/10$ were above 95\% are marked in dark grey.
}\label{tab:comptable_3yr}
\centering
\begin{tabular}[h]{c|c|c|c|c|c|c}

                         &   SE vs.   SC  &   LC vs.   SC  &   LC vs.   SE  &   LE vs.   SC  &   LE vs.   SE  &   LE vs.   LC  \\
\hline\hline
              $\ln{m_1}$ &         0.2246 & \cclrdg      1 & \cclrdg      1 & \cclrdg  0.979 & \cclrdg 0.9984 & \cclrlg  0.597 \\
\hline
$\ln{m_2}$ \& $\ln{m_1}$ &         0.1926 & \cclrdg      1 & \cclrdg      1 & \cclrdg      1 & \cclrdg      1 & \cclrlg 0.6242 \\
\hline
              $\ln{m_2}$ &         0.0838 & \cclrdg      1 & \cclrdg      1 & \cclrdg      1 & \cclrdg      1 &          0.299 \\
\hline
$\ln{D_L}$ \& $\ln{m_1}$ &         0.1956 & \cclrdg      1 & \cclrdg      1 & \cclrdg  0.976 & \cclrdg 0.9954 & \cclrgr 0.7702 \\
\hline
$\ln{D_L}$ \& $\ln{m_2}$ &         0.0796 & \cclrdg      1 & \cclrdg      1 & \cclrdg      1 & \cclrdg      1 & \cclrlg  0.589 \\
\hline
              $\ln{D_L}$ &          0.069 & \cclrdg      1 & \cclrdg 0.9994 & \cclrdg 0.9696 & \cclrdg 0.9318 &         0.4972 \\
\hline
    Spin 1 \& $\ln{m_1}$ & \cclrdg      1 & \cclrdg      1 & \cclrdg      1 & \cclrdg      1 & \cclrdg 0.9996 & \cclrdg      1 \\
\hline
    Spin 1 \& $\ln{m_2}$ & \cclrdg      1 & \cclrdg      1 & \cclrdg      1 & \cclrdg      1 & \cclrdg      1 & \cclrdg      1 \\
\hline
    Spin 1 \& $\ln{D_L}$ & \cclrdg      1 & \cclrdg      1 & \cclrdg      1 & \cclrdg      1 & \cclrdg 0.9814 & \cclrdg      1 \\
\hline
                  Spin 1 & \cclrdg      1 & \cclrdg 0.9984 & \cclrdg      1 & \cclrdg      1 & \cclrdg 0.9728 & \cclrdg      1 \\
\hline
    Spin 2 \& $\ln{m_1}$ & \cclrdg      1 & \cclrdg      1 & \cclrdg      1 & \cclrdg      1 & \cclrdg  0.998 & \cclrdg      1 \\
\hline
    Spin 2 \& $\ln{m_2}$ & \cclrdg      1 & \cclrdg      1 & \cclrdg      1 & \cclrdg      1 & \cclrdg      1 & \cclrdg      1 \\
\hline
    Spin 2 \& $\ln{D_L}$ & \cclrdg      1 & \cclrdg      1 & \cclrdg      1 & \cclrdg      1 & \cclrdg 0.9732 & \cclrdg      1 \\
\hline
    Spin 2 \&     Spin 1 & \cclrdg      1 & \cclrdg 0.9994 & \cclrdg      1 & \cclrdg      1 & \cclrdg 0.9826 & \cclrdg      1 \\
\hline
                  Spin 2 & \cclrdg 0.9998 & \cclrdg 0.9992 & \cclrdg      1 & \cclrdg      1 & \cclrdg 0.9562 & \cclrdg      1 \\
\hline
\end{tabular}
\end{table}

\begin{figure*}
   \centering
   \includegraphics[width=1.0\textwidth]{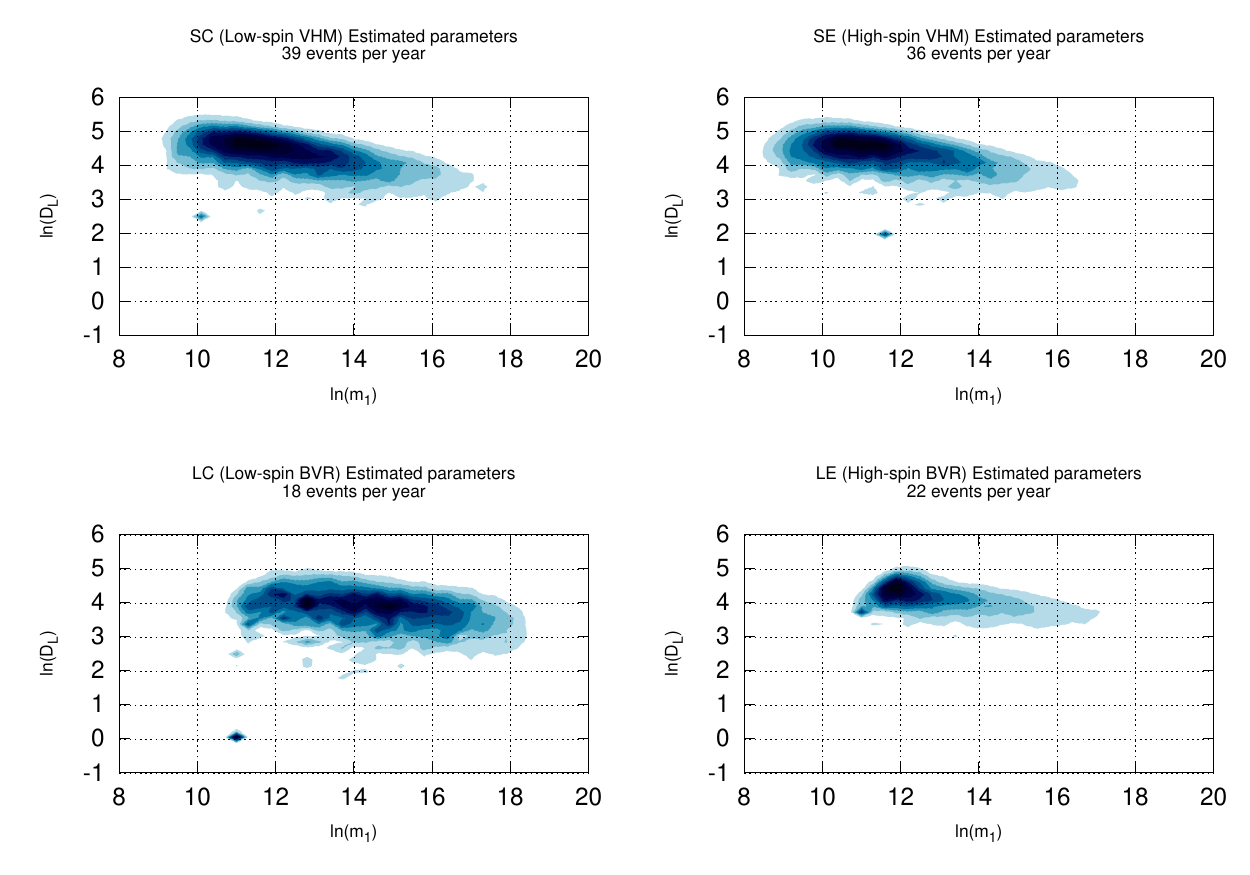}
   \caption{Distributions of estimated $m_1$ and $D_L$ parameters.  With 3 years of sources, the SC and SE models cannot consistently be distinguished, while the LE and LC models can often (in $77\%$ of realisations) be distinguished at $95\%$ confidence. Other 3 year comparisons consistently give $95\%$ or greater confidence levels.}
   \label{fig:ExampleDLm1distributions}
\end{figure*}

\begin{figure*}
   \centering
   \includegraphics[width=1.0\textwidth]{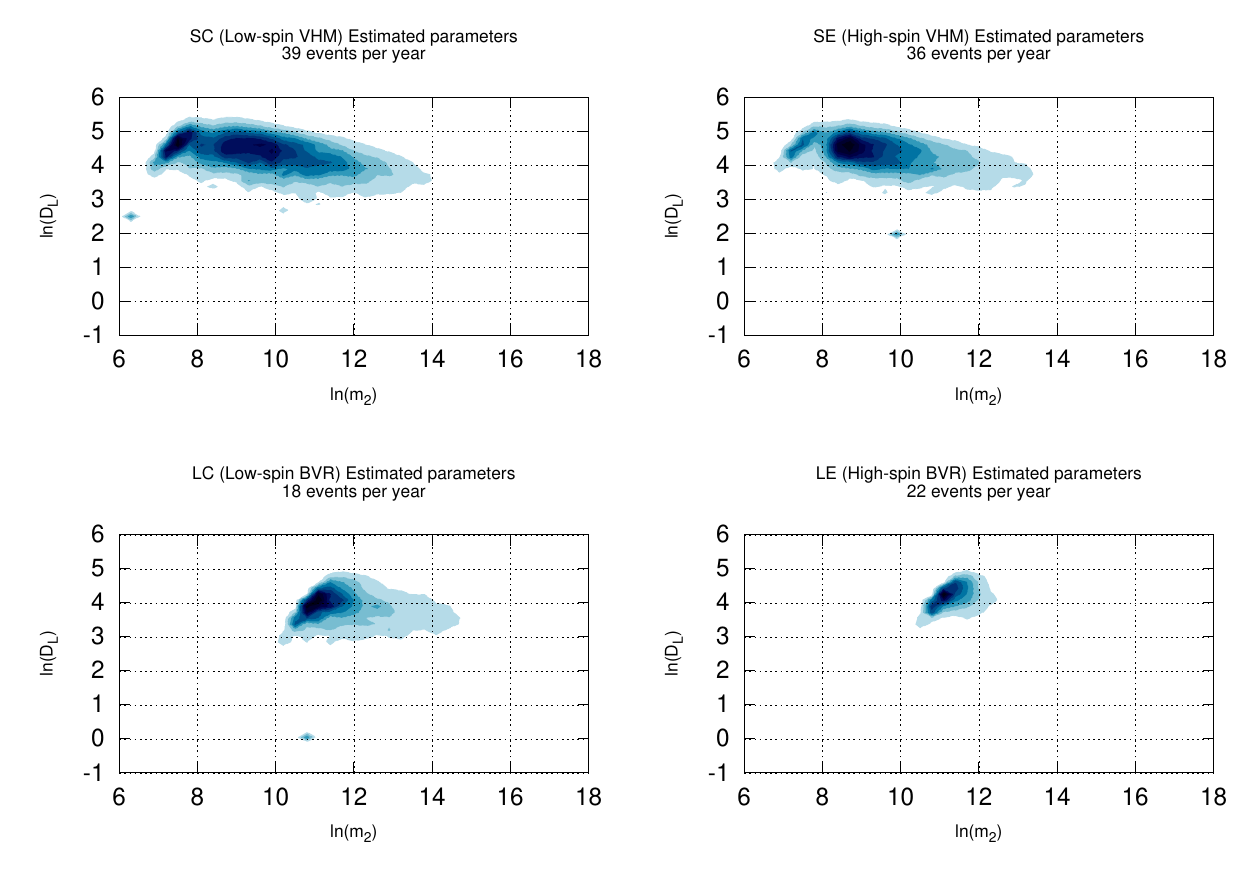}
   \caption{Distributions of estimated $m_2$ and $D_L$ parameters.  With 3 years of sources, the SC and SE models cannot consistently be distinguished, while the LE and LC models can sometimes (in $59\%$ of realisations) be distinguished at $95\%$ confidence. Other 3 year comparisons consistently give $95\%$ or greater confidence levels.}
   \label{fig:ExampleDLm2distributions}
\end{figure*}

Since we are interested as much in the limits of LISA's ability to distinguish between the population models as in whether or not LISA will be able to distinguish between these particular sets of models (which will undoubtedly be superseded with improved versions by the time LISA is operational), Figures \ref{fig:ExampleDLm1distributions} and \ref{fig:ExampleDLm2distributions} show the distributions of $\ln{D_L}$ and $\ln{m_1}$, and of $\ln{D_L}$ and $\ln{m_2}$. The plots serve to illustrate distributions which have varying levels of distinguishability.

\subsubsection{Effects of Parameter Estimation Errors on Model Distinguishability}\label{sub:erroreffects}

In order to assess the significance of the effects of parameter estimation error on the distinguishability of the models, we have also performed model comparison tests with no parameter estimation errors. The results of these tests, for one year of observations, are shown in Table \ref{tab:comptable_noerrs}. The differences between these model distinguishability fractions and those found in Table \ref{tab:comptable_1yr} (which incorporate the parameter estimation errors) are shown in Table \ref{tab:comptable_differences}. These differences show how the fraction of significantly different model realisations decreases when the parameter estimation errors are applied. 

For comparison of the SE and SC models (and to a lesser extent for the LE and LC models), the parameter estimation errors in the spin of the smaller BH significantly decrease the fraction of significantly different model realisations, as anticipated in Section \ref{sec:parmerrors}. The parameter estimation errors in $D_L$ might be expected to have an effect on the distinguishability of some of the model parameter distributions, but it appears that they do not. This is not so surprising when we remember that the source distributions of $D_L$ (Figure \ref{fig:DLdetdistributions}) are fairly smooth on the scale of the parameter estimation errors in $D_L$ (Figure \ref{fig:DLerrdistributions}). The effects of parameter estimation error on the distributions of the spin of the second BH and on $D_L$ can be seen in Figures \ref{fig:spin2detdistributions} and \ref{fig:DLdetdistributions}, respectively. Other than the distributions of spin 2 and $D_L$, the parameter estimation errors have no significant effect on the fraction of confidence levels above $95\%$. Of the distributions considered here, only the spin distributions are sharply peaked enough (compared to the errors) for the parameter estimation errors to have a significant effect on model distinguishability.

In some cases, application of the parameter estimation errors causes the confidence levels to increase slightly. This is due to cases where one of the parameters (e.g., $D_L$) for two models are distributed quite similarly, but the parameter estimation errors are significantly different between the models, due to one of the other parameters (such as the spins) affecting the GW signal having very different distributions. Thus, the source parameter (e.g., $D_L$) distribution can be very similar, while the estimated parameter distribution can be different (due to a larger spread in the distribution of the source with larger parameters). This does not mean that it is possible for the models to be more distinguishable on the whole with errors than without, however, since we are ignoring the parameter distributions (the spins in this example) that are very different and concentrating on just one of the parameter dimensions of the model. When we consider the distinguishability of the models across all of the parameters which affect the signal, we should find that the models are indeed less distinguishable when the parameter estimation errors are applied. This is what we see when we look at Table \ref{tab:comptable_differences}: the parameter estimation errors for each of the model comparisons decreases the fraction of confidence levels above $95\%$ (the positive values in the table) more frequently than it increases them (the negative values in the table).

\begin{table}\footnotesize
\caption{E statistic model comparison results for 1 year of observation with no parameter estimation errors, showing the fraction of confidence levels which were above 95\%. 
%To make the table easier to read, comparisons where $1/2$ of the confidence levels were above 95\% are marked in light grey, comparisons where $3/4$ were above 95\% are marked in medium grey, and those where $9/10$ were above 95\% are marked in dark grey.
}\label{tab:comptable_noerrs}
\centering
\begin{tabular}[!h]{c|c|c|c|c|c|c}

%\begin{tabular}[!h]{c|c|c}
%\hline 
                         &   SE vs.   SC  &   LC vs.   SC  &   LC vs.   SE  &   LE vs.   SC  &   LE vs.   SE  &   LE vs.   LC  \\
\hline\hline
              $\ln{m_1}$ &          0.078 & \cclrdg 0.9188 & \cclrdg 0.9278 & \cclrlg 0.5648 & \cclrlg 0.6006 &         0.1626 \\
\hline
$\ln{m_2}$ \& $\ln{m_1}$ &         0.0858 & \cclrdg 0.9892 & \cclrdg 0.9924 & \cclrdg 0.9708 & \cclrdg 0.9878 &         0.1964 \\
\hline
              $\ln{m_2}$ &         0.0616 & \cclrdg 0.9894 & \cclrdg 0.9914 & \cclrdg 0.9842 & \cclrdg  0.995 &         0.1148 \\
\hline
$\ln{D_L}$ \& $\ln{m_1}$ &         0.0878 & \cclrdg 0.9556 & \cclrdg 0.9536 & \cclrlg 0.5458 & \cclrlg  0.559 &         0.2904 \\
\hline
$\ln{D_L}$ \& $\ln{m_2}$ &         0.0696 & \cclrdg 0.9942 & \cclrdg  0.995 & \cclrdg 0.9672 & \cclrdg 0.9888 &         0.2334 \\
\hline
              $\ln{D_L}$ &         0.0586 & \cclrgr 0.8988 & \cclrgr 0.8486 & \cclrlg 0.5454 &         0.4476 &         0.1816 \\
\hline
    Spin 1 \& $\ln{m_1}$ & \cclrdg      1 & \cclrdg 0.9288 & \cclrdg      1 & \cclrdg      1 & \cclrlg  0.703 & \cclrdg 0.9878 \\
\hline
    Spin 1 \& $\ln{m_2}$ & \cclrdg      1 & \cclrdg 0.9828 & \cclrdg      1 & \cclrdg      1 & \cclrdg 0.9938 & \cclrdg 0.9774 \\
\hline
    Spin 1 \& $\ln{D_L}$ & \cclrdg      1 & \cclrgr 0.8974 & \cclrdg 0.9996 & \cclrdg      1 & \cclrlg 0.6038 & \cclrdg  0.973 \\
\hline
                  Spin 1 & \cclrdg      1 & \cclrgr 0.7842 & \cclrdg 0.9998 & \cclrdg      1 & \cclrlg  0.585 & \cclrdg 0.9846 \\
\hline
    Spin 2 \& $\ln{m_1}$ & \cclrdg  0.999 & \cclrgr 0.8658 & \cclrdg 0.9992 & \cclrdg  0.999 & \cclrlg 0.6548 & \cclrdg  0.981 \\
\hline
    Spin 2 \& $\ln{m_2}$ & \cclrdg 0.9996 & \cclrdg 0.9638 & \cclrdg      1 & \cclrdg      1 & \cclrdg 0.9878 & \cclrdg 0.9734 \\
\hline
    Spin 2 \& $\ln{D_L}$ & \cclrdg 0.9994 & \cclrgr 0.8654 & \cclrdg 0.9992 & \cclrdg 0.9988 & \cclrlg 0.5158 & \cclrdg 0.9718 \\
\hline
    Spin 2 \&     Spin 1 & \cclrdg      1 & \cclrlg 0.7154 & \cclrdg      1 & \cclrdg      1 & \cclrlg 0.6704 & \cclrdg 0.9998 \\
\hline
                  Spin 2 & \cclrdg 0.9998 & \cclrlg 0.7172 & \cclrdg 0.9996 & \cclrdg 0.9996 & \cclrlg 0.5142 & \cclrdg 0.9864 \\
\hline
\end{tabular}
\end{table}

\begin{table}\footnotesize
\caption{The difference between the fraction of confidence levels which were above 95\% without parameter estimation errors and with parameter estimation errors. 
%To make the table easier to read, differences above $0.1$ are marked in light grey, differences above 0.2 are marked in medium grey, and those above 0.3 are marked in dark grey.
}\label{tab:comptable_differences}
\centering
\begin{tabular}[!h]{c|c|c|c|c|c|c}

%\begin{tabular}[!h]{c|c|c}
%\hline 
                         &    SE vs.    SC &    LC vs.    SC &    LC vs.    SE &    LE vs.    SC &    LE vs.    SE &    LE vs.    LC \\
\hline\hline
              $\ln{m_1}$ &          0.0004 &          0.0008 &          0.0004 &         -0.0012 &          0.0014 &         -0.0004 \\
\hline
$\ln{m_2}$ \& $\ln{m_1}$ &         -0.0002 &          0.0002 &         -0.0002 &         -0.0008 &               0 &         -0.0018 \\
\hline
              $\ln{m_2}$ &         -0.0002 &          0.0004 &          0.0004 &               0 &          0.0002 &               0 \\
\hline
$\ln{D_L}$ \& $\ln{m_1}$ &         -0.0016 &          0.0052 &          0.0004 &          0.0162 &          0.0088 &          -0.006 \\
\hline
$\ln{D_L}$ \& $\ln{m_2}$ &          -0.002 &          0.0006 &          0.0004 &         -0.0006 &          0.0012 &         -0.0042 \\
\hline
              $\ln{D_L}$ &          0.0024 &          0.0032 &          0.0064 &          0.0118 &           0.011 &         -0.0018 \\
\hline
    Spin 1 \& $\ln{m_1}$ &               0 &          0.0134 &               0 &               0 &          0.0288 &          0.0024 \\
\hline
    Spin 1 \& $\ln{m_2}$ &               0 &          0.0054 &               0 &               0 &         -0.0026 &          0.0016 \\
\hline
    Spin 1 \& $\ln{D_L}$ &               0 &          0.0062 &          0.0004 &               0 &          0.0834 &          0.0056 \\
\hline
                  Spin 1 &               0 &          0.0002 &          0.0002 &               0 &          0.0764 &          0.0038 \\
\hline
    Spin 2 \& $\ln{m_1}$ & \cclrgr  0.2032 &         -0.0916 &           0.001 &          0.0198 &          -0.038 &           0.089 \\
\hline
    Spin 2 \& $\ln{m_2}$ & \cclrgr  0.2114 &          -0.026 &               0 &          0.0008 &         -0.0066 & \cclrlg  0.1548 \\
\hline
    Spin 2 \& $\ln{D_L}$ & \cclrgr  0.2356 &         -0.0432 &          0.0296 &          0.0366 &          0.0022 & \cclrgr  0.2036 \\
\hline
    Spin 2 \&     Spin 1 &               0 &         -0.1206 &          0.0004 &               0 &          0.0804 &           0.008 \\
\hline
                  Spin 2 & \cclrgr  0.2324 &         -0.1156 &           0.035 &          0.0166 &          0.0202 & \cclrgr  0.2004 \\
\hline
\end{tabular}
\end{table}

\subsection{Summary}

This work makes use of 4 sets of model simulation results \cite[see also][]{LISAPE2009}, each of which uses a different combination of two accretion (`chaotic' - low spins and `efficient' - high spins) and seeding (low mass and high mass) scenarios. The results are labelled `Small Chaotic' (SC), `Small Efficient' (SE), `Large Chaotic' (LC), and `Large Efficient' (LE). We have applied the LISA parameter estimation errors and SNR thresholds to each, allowing us to produce realisations of the LISA estimated parameter distributions predicted by the models. We then compare these estimated parameter distributions using the 1 and 2-Dimensional variants of the K-S test developed in Section \ref{chap:modelcomp}, investigating how the distinguishability of the models varies with the choice of parameter compared, the length of the observation window, and other factors. The results are as follows:

\begin{itemize}
\item All of the models compared can be consistently distinguished with a single year of LISA data, the spin distributions of the chaotic accretion models (SC, LC) being almost trivially distinguishable from those of the efficient accretion models (SE, LE).
\item The parameter estimation errors in the spin of the smaller BH can have appreciable effects on the distinguishability of the models, while those in the masses and the spin of the large BH do not. The parameter estimation errors in $\ln{D_L}$ are too small, compared to the rather smooth model $\ln{D_L}$ distributions, to have a significant effect on the model distinguishability.
\item The distributions of estimated $\ln{m_1}$ and $\ln{D_L}$ for each of the four models illustrate which parameter distributions LISA can distinguish with ease, and which are more difficult to distinguish. For instance, LISA cannot distinguish between these distributions for the SE vs. SC comparison, but can often distinguish between them for the LE vs. LC comparison (given 3 years of observation).
\end{itemize}

The ease with which these models can be distinguished demonstrate the promise of LISA, and point the way forward for MBH population modellers and model comparison testing. It is clear from these results that LISA is uniquely suited to examining and constraining the distribution of the cosmological MBH population. Moreover, since these models are so readily distinguished, it is clear that LISA will also be able to discriminate between significantly more subtle variations between the populations.

Note that the model comparisons performed here made use only of the estimated, or best-fit, parameters associated with sources found in a realisation of the LISA data stream. They do not consider the posterior distribution of likely source parameters associated with each binary whose presence is inferred from a realisation of the LISA data stream.ÊIn reality, we will be able to determine from the LISA waveform that some sources will have better signals (e.g., those with a higher SNR or longer observation time) than others, and the range of binary parameters which could have given rise to these detected events will be more tightly constrained. Thus, some sources will have a larger contribution to our state of knowledge about the population models than others. In order to take these varying contributions into account, Bayesian model selection methods should be employed.

\section*{Acknowledgements}
We would like to acknowledge substantial assistance from Neil Cornish, Marta Volonteri, and Shane Larson. The work of JEP was supported in part by grants and fellowships from Montana NASA EPSCoR, the Montana NASA Space Grant Consortium, and also by the William Hiscock Memorial Scholarship.

%\bibliographystyle{mn}

%\bibliography{mn-jour,MBHPopConstrain2}

\bibliography{MBHPopConstrain2}

%\appendix% put appendix chapters here.
%% \input{appendixWedge}
%\input{appendixFigs}
%\input{appendixCode}
\end{document}